\documentclass[aps,pra,twocolumn,showpacs,superscriptaddress,longbibliography]{revtex4-2}

\usepackage{lineno}

\usepackage[T1]{fontenc}
\usepackage{graphicx}% Include figure files
\usepackage{dcolumn}% Align table columns on decimal point
\usepackage{mathtools} 
\usepackage{bm,amssymb}% bold math
\usepackage{braket}
\usepackage[normalem]{ulem}
\usepackage{xcolor} 
\usepackage[most]{tcolorbox}
\usepackage{booktabs}
\usepackage{diagbox}
\usepackage{soul}
\graphicspath{{figures/}}

\renewcommand\thesection{\arabic{section}}

\makeatletter
\renewcommand\section{\@startsection
  {section}{1}{\z@}%
  {-2.0ex \@plus -0.8ex \@minus -0.2ex}%
  {0.9ex \@plus 0.2ex}%
  {\normalfont\bfseries\raggedright}}
\renewcommand\subsection{\@startsection
  {subsection}{2}{\z@}%
  {-1.6ex \@plus -0.6ex \@minus -0.2ex}%
  {0.6ex \@plus 0.2ex}%
  {\normalfont\bfseries\raggedright}}
\makeatother

\begin{document}

%\linenumbers

\title{Nonlinear optical spectra from Rydberg-mediated photon-photon interactions}

%Probe interaction in microwave-dressed Rydberg electromagnetically induced transparency
%Spectra nonlinearity from Rydberg mediated probe interactions
% Spectra nonlinearity from photon-photon interactions
% "probe interaction" in a title is not understandable

\author{Xinghan Wang}%
\affiliation{ 
Department of Physics and Astronomy, Purdue University, West Lafayette, IN 47907, USA
}%

\author{Yupeng Wang}%
\affiliation{ 
Department of Physics and Astronomy, Purdue University, West Lafayette, IN 47907, USA
}%

   \author{Aishik Panja}

\affiliation{ 
Department of Physics and Astronomy, Purdue University, West Lafayette, IN 47907, USA
}

\author{Qi-Yu Liang}
\affiliation{ 
Department of Physics and Astronomy, Purdue University, West Lafayette, IN 47907, USA
}%
\affiliation{Purdue Quantum Science and Engineering Institute, Purdue University, West Lafayette, IN 47907, USA}

\date{\today}

\begin{abstract}
While Rydberg–Rydberg interactions are essential for quantum nonlinear optics and quantum information processing, their role in microwave and radio-frequency sensing remains poorly understood. Here we experimentally investigate Rydberg interaction-induced nonlinearity in cold-atom Rydberg electromagnetically induced transparency (EIT). In a three-level EIT system, increasing photon-photon interactions produces nonlinear spectral broadening accompanied by resonance shifts, while a microwave-dressed four-level system exhibits pronounced nonlinear broadening without detectable spectral shifts.
Our three-level data can be explained by a conditional superatom model, whereas our four-level observations are surprisingly captured by a simple dephasing model. 
Comparisons with three representative models provide key insights to the role of many-body interactions in Rydberg EIT spectroscopy. Furthermore, our results clarify the conditions under which microwave field characterization can be performed in the nonlinear regime without introducing systematic bias. Our study advances both fundamental understanding of many-body physics and practical development of atomic sensors.
% Furthermore, our results clarify the conditions under which microwave field characterization can be performed in the nonlinear regime without introducing systematic bias, while reducing atomic and photonic shot noise.
%Furthermore, our results point towards conditions for microwave field characterization in the nonlinear regime without introducing measurable systematic errors, while reducing atomic and photonic shot noise.
 %Furthermore, our results show that cold-atom Rydberg sensors can operate in the nonlinear regime, reducing atomic and photonic shot noise without introducing measurable systematic errors. 
%Microwave Rabi frequencies and extracted field amplitudes remain statistically unchanged across probe rates. This holds true for all microwave polarizations. 
%By comparing three representative theoretical models, we find that interaction-induced dephasing provides an effective description of the nonlinear response in the microwave-dressed system, while unconditional mean-field models fail. 
%Ideally, the main text (not including Abstract, Methods, References and Figure legends)  should be limited to 5,000 words. The maximum title length should be 15 words. The abstract — which should be no more than 200 words long and contain no references — should serve both as a general introduction to the topic and as a brief, non-technical summary of the main results and their implications.
\end{abstract}

\maketitle
%\section{Introduction}

Rydberg atoms provide a uniquely versatile platform for engineering strong, tunable interactions in otherwise weakly interacting neutral-atom systems. They couple strongly to each other and to external fields. The former enables controllable atom-atom, atom-photon and even photon-photon interactions. Strong atom–atom interactions have established atom arrays as a leading platform for quantum information processing and quantum simulation~\cite{henriet2020quantum,graham2022multi,muniz2025high,saffman2010quantum}, while atom-photon and Rydberg-mediated effective photon–photon interactions lead to breakthroughs in quantum optics~\cite{chang2014quantum,adams2019rydberg,liang2018observation} and underpin a broad range of optical quantum technologies, such as on-demand single-photon sources~\cite{ornelas2020demand}, deterministic photonic entangling gates~\cite{stolz2022quantum,vaneecloo2022intracavity}, non-classical photonic state generation~\cite{bekenstein2020quantum}, distributed quantum computing~\cite{oh2023distributed} and quantum networking~\cite{grankin2018free}. In the latter case, strong coupling to external fields promises exceptionally sensitive and broadband microwave (MW) and radio frequency (RF) sensing capabilities~\cite{meyer2020assessment,schlossberger2024rydberg,zhang2024rydberg}. 
Rydberg sensors span over six orders of magnitude in frequency, from below one megahertz to terahertz range~\cite{adams2019rydberg,kitching2025atom}, and provide International System of Units (SI) traceable, self-calibrated measurements.

Although the vast majority of MW/RF sensors employ vapor cells, laser-cooled atoms have drawn increasing attention in recent years. Cold atomic systems significantly suppress
Doppler broadening that limits vapor cell sensitivity, offering a path towards quantum-limited performances~\cite{TuSQLelectrometer,duverger2024metrology,jamieson2025continuous,liao2020microwave,zhou2023improving}.
Beyond improved performance in traditional sensing schemes, cold atoms open up new approaches, such as leveraging quantum entanglement~\cite{ocola2024control,chen2024quantum}, coherence~\cite{shu2024eliminating}, and avalanche gain~\cite{nill2024avalanche,wang2025robust}.

However, the role of Rydberg-Rydberg interactions in MW/RF sensing remains vastly underexplored. While these interactions are essential for quantum information processing applications, in sensing contexts they could introduce resonance shifts that lead to systematic errors, and increased decoherence that broadens spectral features. Sensing signals are typically optically read out through electromagnetically induced transparency (EIT). While interaction effects on the resulting transmission spectra are likely to be more pronounced in cold-atom systems, their role in vapor-cell sensors also remains uncharacterized. This gap calls for a deeper understanding of how underlying atomic interactions impact the prospects and limitations of atomic sensors.

Beyond these practical sensing considerations, from a fundamental-physics perspective, it is important to understand how many-body interactions modify the EIT response. At low photon-rate limit, the nonlinear effect can be well captured by an effective field theory~\cite{gullans2016effective}. However, this method is intractable at high photon rates, reflecting the broader challenge of describing strongly interacting many-body systems. To grasp the core physics, we first consider a three-level EIT system without MW/RF coupling. From a mean-field perspective, since van der Waals interactions, characterized by a $C_6$ coefficient, produce level shifts with a definite sign, the resulting spectra are expected to show resonance shifts following the sign of the underlying $C_6$ interaction. However, early experimental observations~\cite{peyronel2012quantum,pritchard2010cooperative} contradict this picture; the nonlinearity appeared as lowered resonance peak heights, while no appreciable shifts were reported.
As such, a model~\cite{petrosyan2011electromagnetically} that accounts for interaction saturation within a blockade radius was introduced and shown to capture the main observed features; however, it still predicts a resonance shift that was not observed experimentally.
As a result, the existing work remains insufficient to establish consensus in the field and does not address the consequences of Rydberg-Rydberg interactions in the presence of MW/RF driving.
% If this model produces no shift for us, it's hard to explain

Here, we experimentally explore the impact of atomic interactions in (MW-dressed) cold-atom Rydberg EIT. First, we apply probe and control laser fields in a three-level system without MW driving. As the photon-photon interaction is increased by raising the probe photon rate, we observe spectral peak broadening similar to prior studies~\cite{peyronel2012quantum,pritchard2010cooperative}. At the same time, we observe a resonance shift that was not reported in these works and is consistent with the predictions of the model mentioned above. Next, we introduce MW driving in the presence of a large bias magnetic field to spectrally separate Zeeman sublevels, thereby isolating an effective four-level system. In this case, we do not observe appreciable spectral shifts when nonlinear broadening becomes comparable to the three-level case where shifts occur. We compare three representative models currently used in the literature to describe the nonlinearity arising from Rydberg-mediated photon–photon interactions. These models predict qualitatively distinct spectral features: one predicts both peak shifts and broadening, another predicts only shifts, and the third predicts only broadening. This qualitative distinction allows us to make meaningful comparisons with our data at the level of characteristic spectral features, rather than detailed parameter choices.

%Finally, at a smaller magnetic field, the presence of multiple MW polarization components couples two Rydberg manifolds, leading to richer spectral structure. Similar to the second case, when nonlinear broadening is evident, no discernible shifts are observed. 
%From another perspective, the MW parameters extracted from nonlinearly broadened spectra statistically agree with those obtained from the linear regime. 

%Our MW-dressed EIT results show that the Rydberg-Rydberg interaction predominantly acts as an additional dephasing mechanism. Consequently, provided interaction-induced shifts remain negligible, a sensor can be operated in the nonlinear regime to achieve higher probe count rates, at the cost of broadened spectral features.

% B field and electrodes are both accepted in sensing.
% Optical pumping can be achieved by sigma^+ probe itself.
% If we don't have optical pumping nor B field, then we will have the same situation as the "quantum-enhanced" paper.

\section*{Results}
\subsection*{Experimental setup and procedure}

We perform Rydberg EIT in a spin-polarized cloud of $^{87}$Rb atoms. The atoms are initially prepared in the ground state $\ket{1} = \ket{5S_{1/2}, F=2, m_F=-2}$, where $F$ and $m_F$ denote the hyperfine and magnetic sublevels, respectively. A weak $\sigma^-$-polarized probe laser couples the ground state to an intermediate state $\ket{2} = \ket{5P_{3/2}, F=3, m_F=-3}$, which is coupled to a Rydberg state $\ket{3}=\ket{3_{-1/2}} =\ket{61S_{1/2},m_j=-1/2}$ via a $\sigma^+$-polarized control laser (Fig.~\ref{fig: EIT nonlinearity}(a)). Additionally, a MW field addresses the $\ket{3} \leftrightarrow \ket{4}=\ket{4_{m_j}}=\ket{61P_{3/2}, m_j }$ transition. Here, $j$ and $m_j$ denote the total angular momentum quantum numbers and the projection of the latter along the quantization axis, respectively.

The atoms are laser-cooled in a magneto-optical trap and transferred to a far-detuned 1064~nm optical dipole trap, following the procedure described in Ref.~\cite{vendeiro2022machine}.  We apply a bias magnetic field $B= 3.7~$G along the probe propagation direction to define the quantization axis and perform optical pumping. The dipole-trapped atomic cloud has a temperature of $14~\mu$K achieved via gray molasses. Doppler broadening is further reduced by applying counter-propagating 780~nm probe and 479~nm control beams. The atomic cloud has a root-mean-square (RMS) radius of $10~\mu$m along the probe propagation direction and an optical depth (OD) of approximately 0.4.

We modulate the dipole trap with a period of 180~$\mu$s and 50\% duty cycle, and probe during the dipole-trap-off time. After 50 modulation cycles, we keep the dipole trap off to allow the atoms to disperse. 
Following a 10~ms time-of-flight, we probe for 2~ms to measure counts without atoms. We average these no-atom counts to obtain a single normalization factor for each spectroscopy scan, which is used to normalize the transmission data. 
We then average the normalized probe transmission over multiple scans. Each spectral data point corresponds to roughly 20~ms of interrogation time. We infer the probe photon rate $R_p$ at the apparatus from the no-atom counts, accounting for the collection and the single photon counting module's (SPCM-AQRH-14-FC) detection efficiency. 

The MW fields are delivered by a broadband horn antenna (FMWAN159-10SF, Fairview Microwave) driven by a signal generator (Anritsu 68369A/NV). Despite using a single antenna, the MW field at the atoms contains all $\sigma^+$, $\sigma^-$ and $\pi$ polarization components due to reflections and scattering from the in-vacuo components, the vacuum chamber and nearby external structures, as is typical in cold-atom experiments~\cite{kurdak2025high}.

% 16.410 entered in the source + 5MHz discrepancy with calculation.

%Following a 10~ms wait time, we probe for 2~ms to accumulate background counts without atoms, which are used to normalize the transmission data for that experimental shot. Each spectral data point corresponds to roughly 20~ms of integration time.  

% 90us per gate x 50 gates x 4 repetition =18ms

 %At large B field, Zeeman sublevels are separated too far so some polarization components of the microwave are not strongly coupled. At low B field, the Zeeman sublevels are close to degenerate so the microwave parameters becomes not resolvable.  

%The probe photon rate is $13$~MHz.%Power measured at the atoms = (2.98+2.38)/2 mW, power after MMF = 0.65 mW, MMF avg count = 1700 for 10 gates each 90us long => MMF count rate = 1.9 Mhz. 60% SPCM efficiency.   Note that sqrt(1700) = 41 ~ 2.4% Measured MMF noise (std dev/ mean) ~3% , so we are not too far off from the poissonian SNR 

\subsection*{Physical model}

We consider three representative models to account for the nonlinearity arising from Rydberg-Rydberg interactions. All models share the concept of superatoms (SAs); namely, all atoms within a blockaded volume $V_b=\frac{4\pi}{3}R_b^3$ collectively behave like a single atom, because the interaction shifts multiple Rydberg excitations out of resonance, preventing doubly- or higher-excited states. Here, $R_b$ denotes the blockade radius and is defined through $C_6^{33}/R_b^6=\Omega_c^2/(2\Gamma_2)$, where $C_6^{33}$ is the van der Waals coefficient of the Rydberg state $\ket{3_{-1/2}}$, $\Omega_c$ is the control Rabi frequency, and $\Gamma_2/(2\pi)=6.1~$MHz is the spontaneous decay rate of the intermediate state $\ket{2}$. These models enforce the blockade condition by restricting the allowed basis states to the ground and singly-excited manifold within each SA. 

Detailed descriptions of the models are provided in "Methods: Model descriptions". Here, we outline the key concepts.
In the \textit{conditional model}~\cite{petrosyan2011electromagnetically}, we spatially coarse-grain the atomic cloud with each pixel corresponding to an SA. The polarizability is determined by two mutually exclusive conditions: if the SA contains a Rydberg excitation, then the polarizability would be the same as that of the system in the absence of control coupling, representing the saturation of interaction (blockade); if the SA is not excited, then atoms outside the SA contribute to a mean-field shift, resulting in a polarizability with a modified control detuning. To implement this model numerically, we consider a 1D SA chain and divide the propagation length $L$ into $L/(2R_b)$ intervals. At each step, we perform Monte Carlo sampling for Rydberg excitation, and evaluate probe field intensity correlation function $g^{(2)}(z)$ and probe Rabi frequency $\Omega_p(z)$ accordingly. The probe transmission $T$ is modeled by averaging $\left|\frac{\Omega_p(z=L)}{\Omega_p(z=0)}\right|^2$ over 50 independent realizations.
% This is equal time g2

Extending this model from the original three-level system to four-level systems requires carefully generalizing the conditional structure of the model in the presence of three interactions with different strengths, characterized by the van der Waals coefficients $C_6^{33}$ and $C_6^{44}$ for $\ket{3}-\ket{3}$ and $\ket{4}-\ket{4}$ interactions, and the dipole-dipole coefficient $C_3^{34}$ for $\ket{3}-\ket{4}$ interactions. In this case, the number and form of the conditioning branches, as well as the associated mean-field contributions, must be reformulated and depend on the hierarchy of the interaction strengths.
We leave the extension of this model to a four-level system to future work and focus instead on contrasting these conceptually distinct models.
%\textbf{All MW EIT data presented here involve only one Zeeman level each in the $\ket{3}$ and $\ket{4}$ manifold, hence  $\ket{3} \equiv \ket{3_{-1/2}}$ and $\ket{4}$ refers to the particular state $\ket{4_{m_j}}$ the microwave is (near) resonant to. (See "Results : Nonlinearity in a four-level MW EIT system".)}

In the \textit{unconditional model}~\cite{wang2025enhanced}, we do not distinguish whether an SA is excited or not, instead calculating the polarizability exclusively with a modified control detuning arising from a mean-field shift. The original work suggests integrating the interaction starting from the interatomic distance to obtain the mean-field results. This leads to very large spectral peak shifts which we do not observe in either our three-level or our four-level systems. To be consistent with other models and the SA framework, we instead set the lower limit to be $R_b$, reflecting that only external SAs outside the blockade volume contribute to the mean-field interactions.

When applied to the four-level system, the mean-field interactions modify the detunings and MW Rabi frequency as $\delta_{2}\rightarrow\delta_2-s_{33}-s_{34}$, $\delta_{3}\rightarrow\delta_3-s_{44}-s_{43}$ and $\Omega_m\rightarrow \Omega_m+2e_{34}$. Here, $s_{33}$ and $s_{44}$ denote the level shifts originating from $C_6^{33}$ and $C_6^{44}$, respectively. Our level $\ket{3}$–$\ket{4}$ interaction exhibits both shift and splitting contributions. Therefore, we fit the pair interaction to the form $C_6^{34}/R^6\pm C_3^{34}/R^3$ ("Methods: Interaction coefficients"). The $C_6^{34}$ term induces shifts $s_{34}$ for level $\ket{3}$ and $s_{43}$ for level $\ket{4}$. The $C_3^{34}$ term contributes to exchange interaction $e_{34}$.
$\delta_2=\Delta_p+\Delta_c$ is the two-photon detuning, $\delta_3=\Delta_p+\Delta_c+\Delta_m$ is the three-photon detuning, $\Omega_m$ is the MW Rabi frequency, and $\Delta_p$, $\Delta_c$ and $\Delta_m$ are probe, control and MW detunings.

Finally, in the \textit{dephasing model}~\cite{TuSQLelectrometer}, we treat the variance of the mean-field interactions as an additional dephasing mechanism, incorporating it as increased Rydberg decoherence rates, rather than modifying the detunings and MW Rabi frequency in the polarizability. Specifically, we modify the jump operators $L_k=\sqrt{\Gamma_k}\ket{k}\bra{k}$ in the Lindbladian by replacing $\Gamma_3\rightarrow\Gamma_3+\sqrt{\text{Var}_{33}}+\sqrt{\text{Var}^s_{34}}+\sqrt{\text{Var}^e_{34}}$ and $\Gamma_4\rightarrow\Gamma_4+\sqrt{\text{Var}_{44}}+\sqrt{\text{Var}^s_{43}}+\sqrt{\text{Var}^e_{34}}$, where $\text{Var}_{33}$, $\text{Var}_{44}$,  $\text{Var}^s_{34}$, $\text{Var}^s_{43}$ and $\text{Var}^e_{34}$ are the variances associated with $s_{33}$, $s_{44}$, $s_{34}$, $s_{43}$ and $e_{34}$, respectively.

For all models, the probe-transition coherence $\rho_{21}$ is calculated from the master equation
\begin{equation}
   \dot{\rho}=\frac{i}{\hbar}[\rho,H] - \sum_{k}\frac{1}{2}\left(  L_k^\dagger L_k\rho  + \rho L_k^\dagger L_k  -2L_k\rho L_k^\dagger \right)
   \label{eq:Liouvillian operator}
\end{equation}
with
\begin{equation}
    \begin{aligned}
    H = & \hbar \bigg( - \Delta_p\ket{2}\bra{2} - \delta_2\ket{3}\bra{3} - \delta_3\ket{4}\bra{4} \\& + \frac{ \Omega_p}{2}\ket{1}\bra{2} + \frac{\Omega_c}{2}\ket{2}\bra{3} + \frac{\Omega_m}{2}\ket{3}\bra{4} + \text{h.c.} \bigg)
    \end{aligned}
\label{eq:H 2-level}
\end{equation}
$L_2=\sqrt{\Gamma_2}\ket{1}\bra{2}$, together with $L_k=\sqrt{\Gamma_k}\ket{k}\bra{k}$, $k\in\{3,4\}$, are the only Lindbladian terms considered. Both the 61S$_{1/2}$ and 61P$_{3/2}$ Rydberg states have lifetime longer than $100~\mu$s at room temperature, resulting in negligible decay rates compared to other decoherence processes.
%Rydberg lifetime is at kilohertz level.
%At 300K, for 61S , it is 1.53 kHz (104 us) and 61P3/2, it is 1.15 kHz (139 us)

The steady-state solution of the density matrix $\rho$ is associated with the probe transmission $T$ via:
\begin{equation}
    T=\exp\left[\text{OD Im}(\rho_{21})\frac{\Gamma_2}{\Omega_p}\right]
    \label{eq: transmission}
\end{equation}
Although $\Omega_p$ explicitly appears in the equation, the spectra exhibit virtually no dependence on $\Omega_p$ in the absence of Rydberg-Rydberg interaction and in the weak-probe limit with respect to single-atom saturation. In this case, the steady-state solution for $\rho_{21}$ is proportional to $\Omega_p$, which cancels out in the normalized probe transmission. For the probe Rabi frequency relevant to our experiments, we operate in this weak-probe limit.
Therefore, the observed spectral dependence on $\Omega_p$ arises from Rydberg-mediated nonlinearity, rather than from single-atom EIT physics.
%We numerically verify that at our highest probe Rabi frequency $\Omega_p = 2\pi\times0.7~\text{MHz}$, doubling $\Omega_p$ changes the spectra by less than 2\% when atomic interactions are neglected. Therefore, any observed spectral dependence on $\Omega_p$ arises from interaction-induced nonlinearity, rather than from single-atom EIT physics.
%where $\Omega_p$ denotes the probe Rabi frequency, and $\Gamma_2=2\pi\times6.1~$MHz is the spontaneous decay rate of the intermediate state.
%We simulate the probe transmission ($T$) as a function of control detuning ($\Delta_c$), with the probe kept resonant with the Zeeman-shifted $\ket{1}\leftrightarrow\ket{2}$ transition, resulting in a low baseline transmission. Under MW dressing, the resulting spectrum can be understood as EIT with these dressed states as third states in the ladder, rather than just $\ket{3_{-1/2}}$. As each MW-dressed state is brought into EIT resonance by scanning $\Delta_c$, a peak in probe transmission is expected.  

\subsection*{Nonlinearity in a three-level system}
\begin{figure}[ht]
\centering
\includegraphics[width=0.45\textwidth]{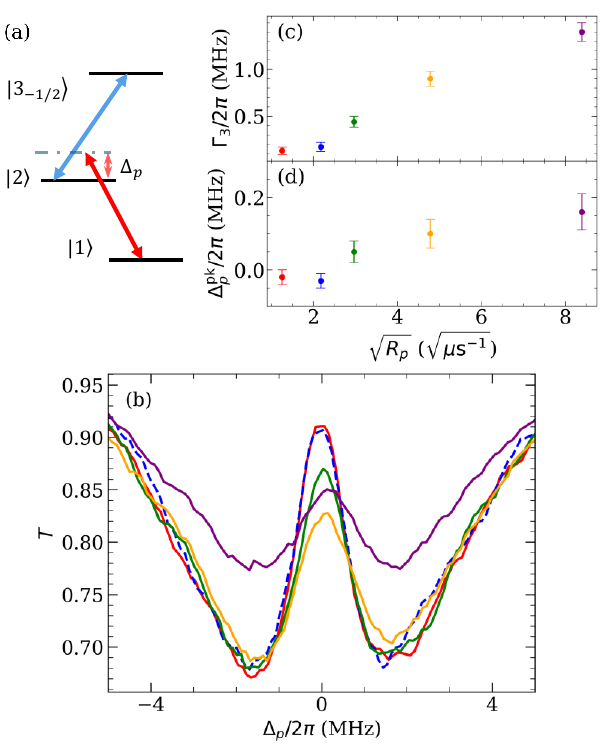}
\caption{Nonlinearity in a three-level system. (a) Level scheme. (b) Measured EIT spectra vs probe detuning. For clarity, we smooth the data using a moving average of window size 9 (see Fig.~\ref{fig: EIT nonlinearity models} for the raw data). The probe rate is $R_p=\{1.6,4.8,8.8,22.9,70.5\}~\mu\text{s}^{-1}$ for the \{red, blue, green, orange, purple\} colors. This color coding is kept the same as in panel (c,d). (c) $\Gamma_3$ is extracted by fitting the linear model to the raw data. (d) The peak position $\Delta_p^{pk}$ is determined by Gaussian fits to the raw data in the central region. 
}
\label{fig: EIT nonlinearity}
\end{figure}
% The drastic probe rate change comes from recalibration of ND filters. We chose only to increase by a factor fo 2 for the highest rate.
%% The language is "fit model to data", not "fit data to model"

The contribution from nonlinear interactions can be systematically varied by the atomic density and the probe Rabi frequency. Increasing the atomic density, and thereby the atom number within a fixed atom-photon interaction volume, improves the signal-to-noise ratio by reducing atomic shot noise, while increasing the probe power reduces photonic shot noise; however, both enhancements also increase nonlinear interactions. This raises the question whether sensing can be performed in the presence of nonlinear effects and whether such effects introduce systematic errors. 

\begin{figure*}[ht]
\centering
\includegraphics[width=0.9\textwidth]{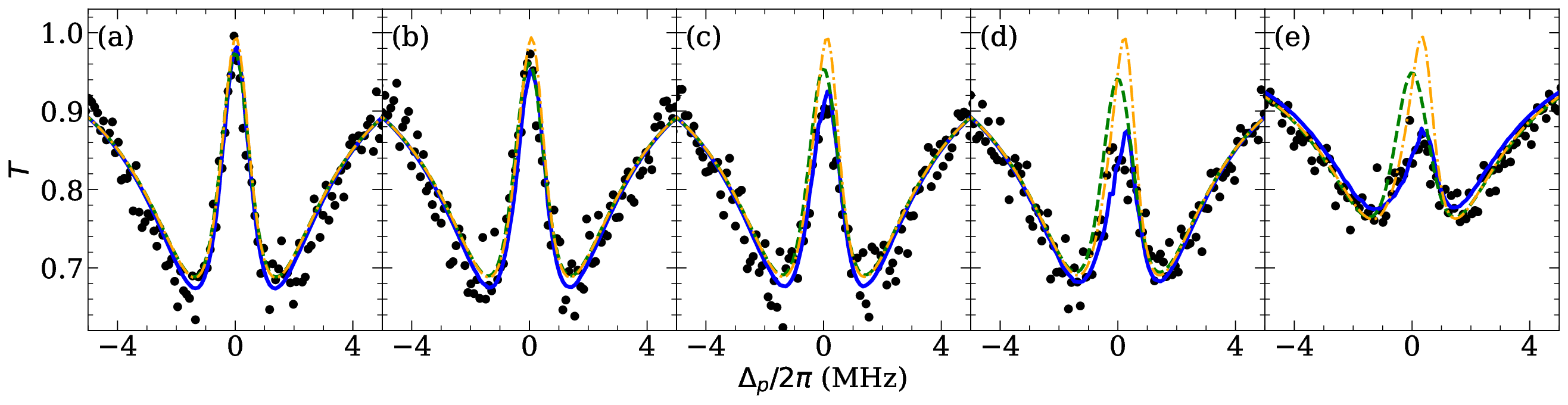}
\caption{Comparison of the three models (conditional model: blue solid line; dephasing model: green dashed line; unconditional model: orange dot-dashed line) with our three-level EIT data (same dataset as in Fig.~\ref{fig: EIT nonlinearity}, shown as black dots). Panels (a–e) correspond to increasing probe rate.}
\label{fig: EIT nonlinearity models}
\end{figure*}

We first investigate a three-level EIT system without MW coupling. As we increase the probe photon rate, the dominant effect is resonant peak height reduction (Fig.~\ref{fig: EIT nonlinearity}(b)), consistent with prior experimental studies~\cite{peyronel2012quantum,pritchard2010cooperative}. To quantify this phenomenon, we fit the \textit{linear model} (Eqs.~\ref{eq:Liouvillian operator} and \ref{eq:H 2-level} with no interaction-induced substitutions and $\Omega_m=0$)
%~\footnote{We numerically solve the master equation which includes possible single-atom saturation, which is insignificant at our probe rate. This model does not include Rydberg interactions, thus dubbed "linear mode" here.} 
to the data (See "Methods: Spectral fits" for details of the parameters) and plot the fitted $\Gamma_3$ values vs the square root of the probe rate (Fig.~\ref{fig: EIT nonlinearity}(c)). At the lowest two rates, the effects of Rydberg-Rydberg interactions are negligible, with $\Gamma_3$ showing no rate dependence. As the rate further increases, $\Gamma_3$ increases. We attribute the OD reduction at the highest rate to atom loss from the probing region due to momentum kicks.
% The language is "fit model to data", not "fit data to model"

Importantly, we observe a resonant blue shift up to 0.2~MHz (Fig.~\ref{fig: EIT nonlinearity}(d)) along with peak height reduction, which increases with probe rate similarly to $\Gamma_3$. This shift is expected from the mean-field perspective, in which the Rydberg interaction between S-state atoms shifts the level upward, resulting in a smaller effective two-photon detuning, thereby blue-shifting the resonant peaks. To ensure that the observed shift is not caused by uncontrolled parameters, we acquire spectroscopic data at different probe Rabi frequencies in randomized order and repeat the entire acquisition sequence multiple times using newly randomized orders.

%Additionally, at the highest $\Omega_p$ in Fig.~\ref{fig: EIT nonlinearity models}(d), nonlinear effects are not solely due to Rydberg interactions: as the scattering rate increases, radiation-pressure-induced momentum kicks lead to atom loss from the probing region, and the optical response deviates from linearity due to saturation effects. These additional nonlinear effects may affect the extracted Rydberg deocherence $\Gamma_3$, but 

%Within our experimental control and resolution (see Supplementary Part 2), no appreciable spectral shift is detected. These behaviors are consistent with previous findings~\cite{peyronel2012quantum,pritchard2010cooperative}.
%The conditional SA model and dephasing model both account for the reduction in EIT peak height, capturing the main features of our data. The conditional SA model further indicates peak shifts of $\lesssim100~\text{kHz}$, which we cannot discern experimentally.
%By contrast, the unconditional SA model predicts no peak height reduction while predicting a MHz-level spectral shift, which clearly contradicts our observations.

To compare our data with the three models, we integrate the interaction contribution from $R_b$ to infinity. Because the van der Waals interaction decays rapidly with distance, the precise choice of the upper integration limit is insignificant. We approximate our medium as homogeneous with an atomic number density of $\varrho=2\times10^{10}~\text{cm}^{-3}$ and a length of $L=64.8~\mu$m such that $L/(2R_b)=4$ is an integer for the Monte Carlo simulation. This length $L$ roughly corresponds to 2.5 times the Gaussian effective length ($\sqrt{2\pi}\times$ RMS length) extracted from absorption imaging. The remaining input parameters to these models are set to the fitted values of the linear model obtained from fits to the lowest-rate data. In the conditional and dephasing model, we include a  baseline Rydberg dephasing rate $\Gamma_3/(2\pi)=20~\text{kHz}$, roughly accounting for Doppler broadening and ambient electric-field inhomogeneity. In the conditional model, this is the total dephasing rate, whereas in the dephasing model it serves as a background to which interaction-induced dephasing is added. In the unconditional model, $\Gamma_3/(2\pi)$ is empirically set to 200~kHz such that the model prediction is consistent with our measurement at the lowest rate.

%The input parameters to these models include the optical depth OD, control Rabi frequency $\Omega_c$, blockade radius $R_b$ and the average number of atoms within a blockade volume $N_{SA}$. Among these parameters, OD and $\Omega_c$, and hence $R_b$, are directly extracted from spectroscopy by scanning the probe detuning $\Delta_p$. $N_{SA}$ can in principle be inferred from the atomic density. However, because the models assume a homogeneous atomic cloud, whereas the experimental cloud has a Gaussian density profile, there is no clear prescription for defining an effective density. As a result, $N_{SA}$ is treated as the sole free parameter and is varied to simultaneously fit all spectra for each model.

As shown in Fig.~\ref{fig: EIT nonlinearity models}, the unconditional model only predicts peak shifts, the dephasing model only predicts peak height reduction, whereas the conditional model predicts both. Only the conditional model adequately explains our observations.   Note that although both the conditional and dephasing model predict peak height reduction, the underlying mechanisms differ starkly. 
In the conditional model, the peak height reduction arises from stochastic conditioning of the response. When a Rydberg excitation is present, the response is projected onto a no-control form, thereby making no contribution to the resonant peak at that step of the Monte Carlo sampling.
Otherwise, it remains close to the usual EIT behavior with a small interaction shift. In contrast, in the dephasing model, the origin of dephasing lies in the random spatial distribution of atoms: atoms at different separations induce different interaction strengths, which introduces a distribution of level shifts and thus dephases the collective response. If the atoms are instead arranged in a lattice structure with well-defined spacings, this positional randomness would be strongly suppressed, and the corresponding dephasing would be minimal.

\subsection*{Nonlinearity in a four-level MW EIT system}

\begin{figure*}[ht]
\centering
\includegraphics[width=0.75\textwidth]{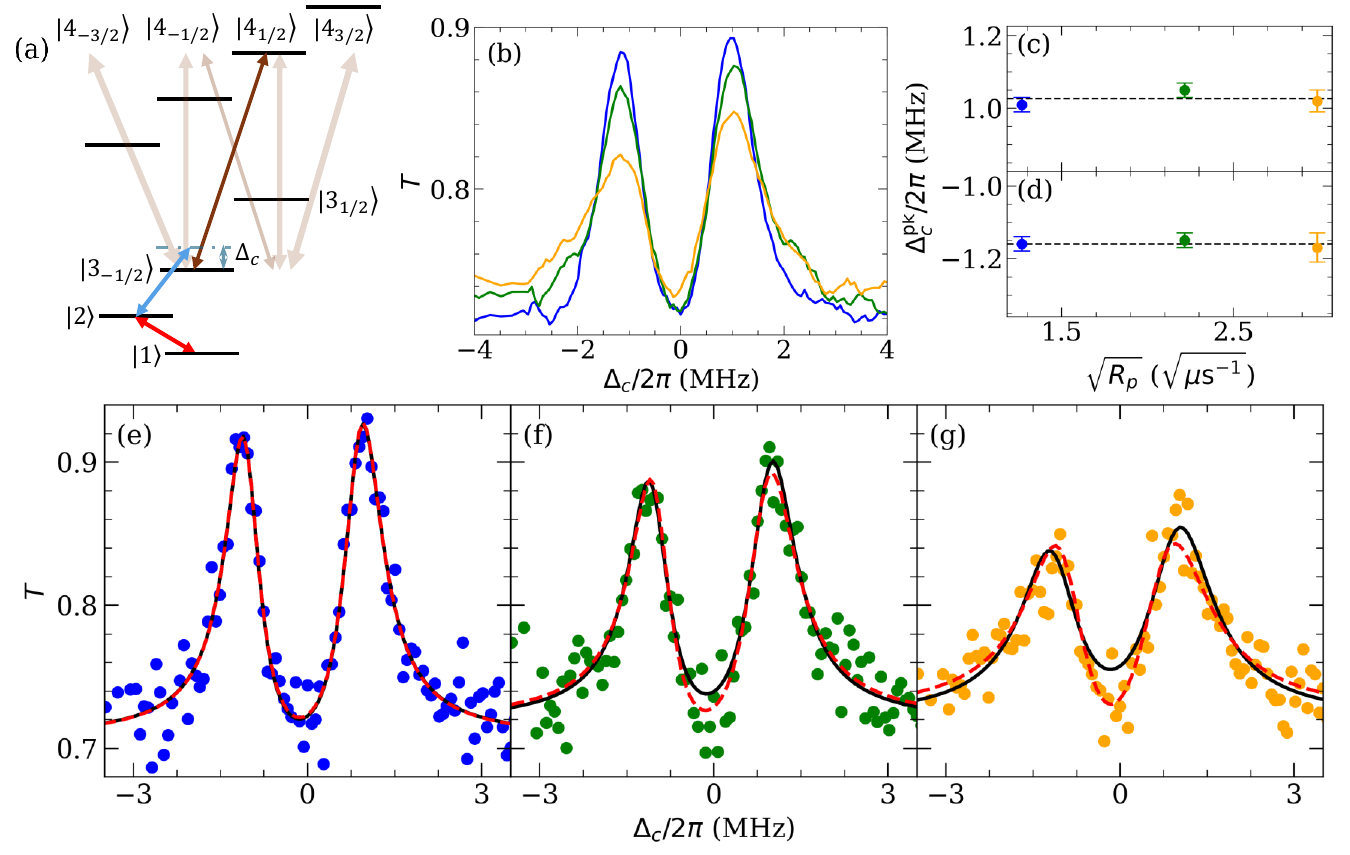}
\caption{Nonlinearity in a four-level system. (a) Level scheme. (b) Measured MW EIT spectra vs control detuning. For clarity, we smooth the data using a moving average of window size 9 (see the dots in panel (e-g) for the raw data). The probe rate is $R_p=\{1.6,4.9,8.9\}~\mu\text{s}^{-1}$ for the \{blue, green, orange\} colors. The slight imbalance between the two peaks results from a small MW detuning $\Delta_m/(2\pi)=0.15~$MHz, which does not affect the extracted values of $\Omega_m$ within statistical uncertainties. The peak positions $\Delta_c^{pk}$ of the right and left peaks are determined by Gaussian fits to the raw data in the vicinity of each peak and plotted in panel (c) and (d). In panels (e-g), the red dashed and black solid lines are fit results from the linear model.  The red dashed lines have independent $\Gamma_3$ and $\Gamma_4$, with fitted values $\Gamma_3/(2\pi)=\{0.17(8),0.6(1),1.3(1)\}~$MHz and $\Gamma_4/(2\pi)=\{0.21(7),0.11(8),0.10(9)\}$~MHz. The black solid lines enforces $\Gamma_3=\Gamma_4$, with fitted values $\Gamma_{3,4}/(2\pi)\ = \{0.19(2),0.32(3),0.68(5)\}$~MHz.}
\label{fig: EIT nonlinearity 2}
\end{figure*}

In the presence of MW, we apply a large magnetic field $B = 15.7$~G and a sufficiently weak MW field, resonant with the Zeeman-shifted $\sigma^+$ transition, such that only a single Zeeman level, $\ket{4_{1/2}}$, is mixed with $\ket{3_{-1/2}}$ (Fig.~\ref{fig: EIT nonlinearity 2}(a)), thereby isolating a four-level subsystem. 
This configuration provides a classic level structure for SI-traceable, self-calibrated MW measurements. When scanning the control detuning $\Delta_c$, while keeping the probe and MW fields on resonance, two resonant peaks arise from the probe absorption baseline. In such measurements, the splitting of the peaks, dubbed Autler-Townes (AT) splitting ($\Delta_{\text{AT}}$), is proportional to the electric field amplitude of the MW ($E_m$) through Planck's constant ($\hbar$) and the dipole moment of the Rydberg transition ($\wp_{34}$): 
\begin{equation}
    E_m=\frac{\hbar}{\wp_{34}}\Delta_{\text{AT}}
\end{equation}
Here, we are interested in how Rydberg-mediated nonlinearity affects this splitting and the spectra.

\begin{figure*}[ht]
\centering
\includegraphics[width=0.7\textwidth]{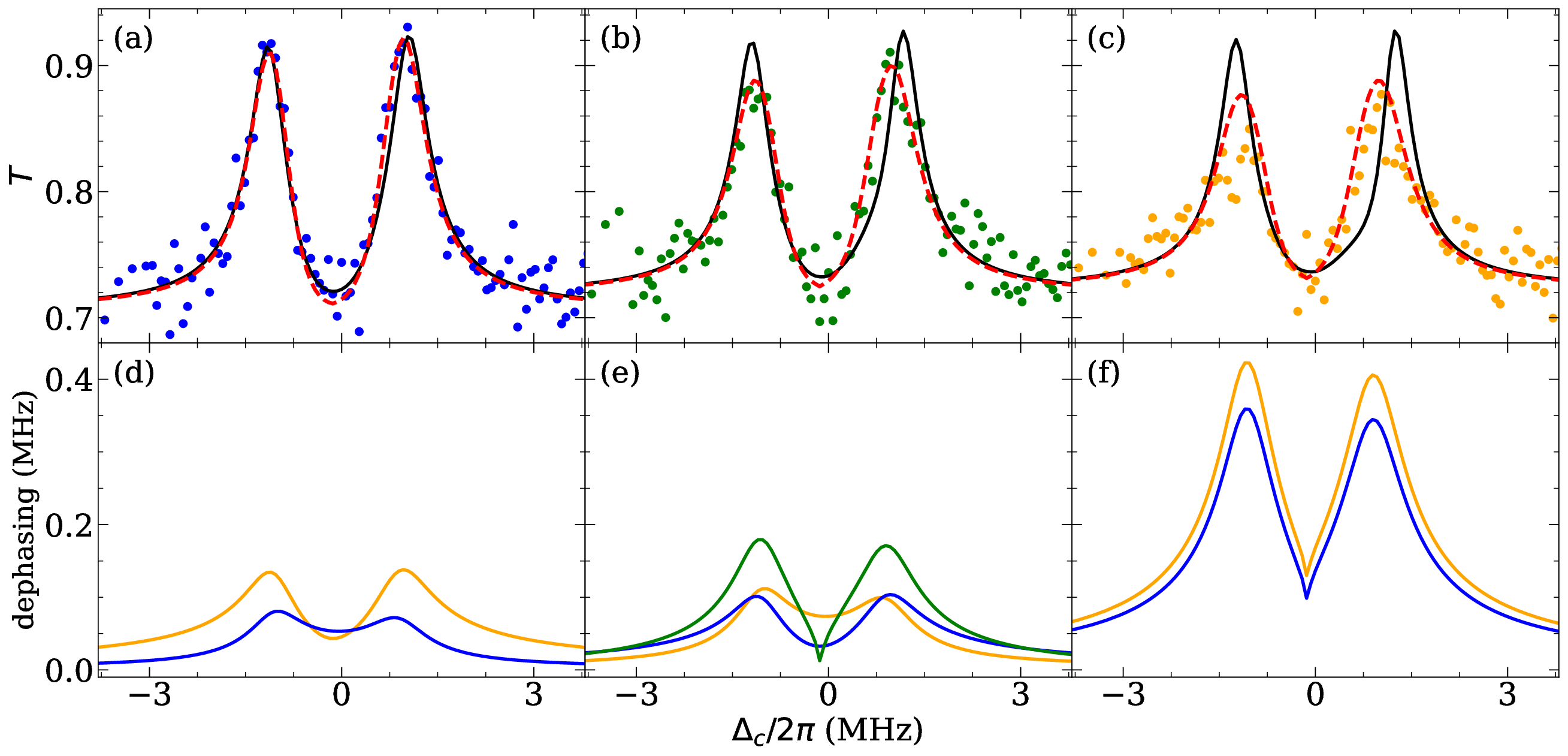}
\caption{Comparison of the unconditional model (black solid lines) and the dephasing model (red dashed lines) with our four-level MW EIT data (same dataset as in Fig.~\ref{fig: EIT nonlinearity 2}, shown as dots.)  Panel (d-f) show interaction-induced Rydberg dephasing in the dephasing model for the highest probe rate $R_p=8.9~\mu s^{-1}$, used for the red dashed curve in panel (c). (d) Contribution  from the $C_6^{33}$ ($\sqrt{\text{Var}_{33}}/(2\pi)$, orange) and $C_6^{44}$ ($\sqrt{\text{Var}_{44}}/(2\pi)$, blue) terms. (e) Contribution from the $C_6^{34}$ term to level $\ket{3}$ ($\sqrt{\text{Var}^s_{34}}/(2\pi)$, orange), from  the $C_6^{34}$ term to level $\ket{4}$ ($\sqrt{\text{Var}^s_{43}}/(2\pi)$, blue) and from the $C_3^{34}$ term to both level $\ket{3}$ and $\ket{4}$ ($\sqrt{\text{Var}^e_{34}}/(2\pi)$, green). (f) Total interaction-induced Rydberg dephasing for level $\ket{3}$ (orange) and $\ket{4}$ (blue). }
\label{fig:  dephasing spectrum}
\end{figure*}

As illustrated in Fig.~\ref{fig: EIT nonlinearity 2}, the peak height reduction sets in clearly at a lower probe rate than in the three-level system. We do not observe any spectral peak shift as a function of probe rate, even though the Rydberg dephasing effect is evident. As a result, the nonlinear spectra can be used for MW detection at the cost of peak broadening. 

The MW field electric field amplitude obtained from the Gaussian-fitted peak positions (Fig.~\ref{fig: EIT nonlinearity 2}(c,d)) is $1.30(2)$~mV/cm, corresponding to a Rabi frequency $\Omega_m/(2\pi)=2.11(4)~$MHz. Alternatively, we can extract these values by fitting the linear model (Eqs.~\ref{eq:Liouvillian operator} and \ref{eq:H 2-level} with no interaction-induced substitutions). For instance, the fit results of the data in Fig.~\ref{fig: EIT nonlinearity 2}(e) are $\{\Delta_m,\Omega_m,\Omega_c,\Gamma_3,\Gamma_4\}/(2\pi)=\{0.15(6),2.04(2),2.76(6),0.17(8),0.21(7)\}$~MHz and OD = 0.347(4). This $\Omega_m$ is 3\% smaller than the value from the AT splitting. In fact, $\Omega_m$ obtained in this manner is systematically smaller than those extracted from the AT splitting. This discrepancy is likely caused by insufficient splitting of the peaks as compared to their widths (see Supplementary Part 1 and Ref.~\cite{holloway2017electric}). When the two peaks are relatively close, the lineshape is skewed, with the inner edges sharper than the outer edges. As such, the peak splitting does not perfectly match the MW Rabi frequency $\Omega_m$. This behavior is tied to underlying EIT response rather than a simple peak-overlap effect. Consequently, alternative peak-finding methods, such as double Gaussian fits, do not remedy this discrepancy.
%Ref.~\cite{holloway2017electric} points out that when $\Omega_m<3\Omega_c^2/\Gamma_2$, the AT splitting deviates from the true MW Rabi frequency. In the regime we operate, the deviation is positive, meaning that the AT splitting overestimates the MW Rabi frequency, consistent with our results.

The distortion of the lineshape does not solely depend on the peak splitting relative to the peak width; $\Gamma_3$ and $\Gamma_4$ contribute to the lineshape in distinct ways.
Increasing $\Gamma_4$ raises the probe transmission between the two resonant peaks. However, as we increase the probe rate, the experimental transmission near $\Delta_c=0$ remains low. Consequently, fitting the linear model consistently returns a small $\Gamma_4$, even showing a trend of decreasing $\Gamma_4$ as the probe rate increases. This trend is opposite to physical expectations, and the resulting large $\Gamma_3$ also suppresses the peak-height imbalance that is clearly visible in the data. These spectral effects become evident by comparing two fitting approaches: one allowing $\Gamma_3$ and $\Gamma_4$ to vary independently (red dashed lines in Fig.~\ref{fig: EIT nonlinearity 2}(e-g)), and one constraining them to be equal (black solid lines in Fig.~\ref{fig: EIT nonlinearity 2}(e-g)). At high probe rates, allowing $\Gamma_4$ to vary freely forces it to smaller values to remain consistent with the low probe transmission near $\Delta_c=0$, whereas constraining $\Gamma_4=\Gamma_3$ fails to reproduce this feature. These behaviors signal a limitation of the linear model with probe-rate-dependent Rydberg decoherence. Nevertheless, both fitting approaches yield statistically identical MW Rabi frequencies until nonlinearity-induced $\Gamma_{3,4}$ significantly broaden the EIT linewidth ($\Gamma_3+\Gamma_4\gtrsim\Omega_c^2/\Gamma_2$), demonstrating robustness against different fitting choices. 
When $\Gamma_{3,4}$ become relatively large, the MW Rabi frequency extracted from model fits to high-rate data can show a slight deviation relative to that obtained from low-rate data.
This deviation reflects model-dependent lineshape distortion, as none of the parameter choices fully reproduce the observed spectra.

%In this regime, the fitted MW Rabi frequency from high-rate data can show a slight deviation relative to that from the low-rate counterpart caused by the bias in the effective peak splitting due to $\Gamma_{3,4}$ lineshape distortion.  

To characterize the full 3D MW field, we vary its frequency and selectively bring the $\sigma^+$-, $\pi$-, or $\sigma^-$-polarized components on resonance. As such, we perform three separate sets of spectroscopic measurements to extract the MW electric field amplitude of the $\{\sigma^+, \pi, \sigma^-\}$ polarization (data for $\pi$ and $\sigma^-$ polarization presented in Supplementary Part 2). The extracted field amplitudes are $\{1.30(2),1.19(1),0.74(1)\}$~mV/cm from the AT splitting and $\{1.25(1),1.13(1),0.67(1)\}$~mV/cm from the linear model fits. Since the MW source power and the experimental setup are otherwise identical, these three polarization components simultaneously exist at the atom position.
 Furthermore, we observe no meaningful difference in probe-rate dependence among the spectra of the three polarizations, despite their different interaction coefficients. We verify that our model predictions are insensitive to the values of the coefficients.
% Across all analysis methods and all polarizations, we do not observe any statistically significant probe rate dependence of MW Rabi frequency
% It's hard not to repeat the "regime" sentence again.
% Perhaps deleting it is fine because I have explained the fits rate-independence. Split rate dependence is bovious from the peak positions
% Different polarizations are addressed in the next sentence.

Since no spectral shifts are observed, the unconditional model (black solid lines in Fig.~\ref{fig:  dephasing spectrum}(a-c)) clearly deviates from our data. Surprisingly,
the dephasing model (red dashed lines in Fig.~\ref{fig:  dephasing spectrum}(a-c)), while insufficient to explain our three-level EIT data, captures the observed features here. Moreover, it predicts detuning-dependent Rydberg decoherence (Fig.~\ref{fig:  dephasing spectrum}(d-f)) such that it increases $\Gamma_3$ and $\Gamma_4$ to comparable values with increasing probe rate, while still capturing the low probe transmission between the two peaks, thereby resolving the apparent inconsistency in the linear-model fits.
If this model is indeed the correct underlying description, it will accurately capture the full lineshape and enable reliable fitting in the deep nonlinear regime. Moreover, another widely used approach to extract the MW Rabi frequency fixes all fields at resonance and performs an atomic superheterodyne measurement~\cite{jing2020atomic,gordon2019weak,prajapati2021enhancement}, which may likewise require corrections for Rydberg-induced nonlinearity.

%Although it is tempting to combine the unconditional model with the dephasing model to account for both shifts and dephasing, such a combination would introduce nontrivial new features in the four-level system, such as kinks. We leave it for future work to evaluate whether this is reasonable.

%We emphasize that the resulting $\Gamma_k$ is detuning-dependent and can show clear differences from a uniformly increased $\Gamma_k$ in the linear model when comparing spectroscopic responses.

%Within this configuration, the fit parameters include the $\sigma^+$-polarized microwave Rabi frequency $\Omega_{MW}^+$ and detuning $\Delta_{MW}^+$, together with the optical depth OD. The decoherence rates in the absence of Rydberg–Rydberg interactions are fixed to $\Gamma_3=\Gamma_4=2\pi\times100~\text{kHz}$, accounting for Doppler broadening as well as ambient electric-field inhomogeneity and drift. This value of $\Gamma_3$ is consistent with fits to three-level EIT spectra obtained by scanning the probe detuning. The control Rabi frequency $\Omega_c$ is likewise fixed using values extracted from the same spectra.

%Dephasing , $\Gamma_{3,deph} = \sqrt{Var_{33}} + \sqrt{Var_{34}} , \Gamma_{4,deph} = \sqrt{Var_{44}} + \sqrt{Var_{34}}$

\section*{Discussion}
In conclusion, we have taken the initial steps to experimentally investigate the nonlinear three-level and four-level EIT spectroscopy arising from Rydberg-Rydberg interactions. In the three-level EIT system, to the best of our knowledge, we are the first to observe interaction-induced peak shifts. The four-level system is particularly relevant in the context of cold-atom MW sensing. The onset of nonlinearity occurs at significantly lower probe rates in the presence of MW fields, manifesting as enhanced Rydberg decoherence, without discernible spectral peak shifts. This implies that optimal sensing performance may be achieved in a nonlinear regime, trading peak broadening for reduced photonic and atomic shot noise.
Since AT splitting determines MW Rabi frequency solely from spectral peak positions, the extracted values are robust against the nonlinearity. For methods sensitive to lineshapes, moderate broadening does not compromise accuracy; however, a model that properly incorporates lineshape distortion may be required to operate reliably in the deep nonlinear regime.

 From the many-body physics point of view, at present, there is no clear consensus in the literature. We compare three representative models, which make qualitatively different predictions. The conditional model is the only one that captures both peak shifts and reductions in peak height when photon-photon interactions increase, which agrees with our three-level EIT observations. In addition, its underlying physical picture is well motivated.
We expect that, when properly extended to the four-level system, the conditional model has the potential to account for the full set of experimental observations. 

Nevertheless, to our surprise, the dephasing model agrees with our four-level data, particularly capturing the spectroscopic features through its detuning-dependent Rydberg dephasing. It would therefore be valuable to identify the deeper physical reasons that justify this agreement and to delineate the conditions under which the model remains valid. Compared with the conditional model, it has the advantage of conceptual and computational simplicity.

In the future, we will apply a small bias magnetic field such that all the Rydberg Zeeman levels are mixed. In that case, we can extract the electric field amplitudes of all three MW polarizations from a single spectroscopic measurement, without varying the MW frequency or other experimental conditions to individually bring each polarization component on resonance. This approach is also sensitive to the relative phases between different polarization components. Based on our observations in the isolated four-level system, we anticipate that operating in moderate nonlinear regime could increase probe counts without introducing systematic bias, though this remains to be verified experimentally.
% We do not the scope of claims. We still need more understanding.

% Our MW-dressed EIT results show that the Rydberg-Rydberg interaction predominantly acts as an additional dephasing mechanism. Consequently, a sensor can, and perhaps should in certain circumstances, be operated in the nonlinear regime to achieve higher probe count rates, at the cost of broadened spectral features.
\section*{Methods}
\subsection*{Spectral fits}
All fits are performed using the \texttt{curve\_fit} function of \texttt{scipy.optimize} python package applied to the linear model,
 where Rydberg-mediated interactions would appear as probe-rate-dependent $\Gamma_{3,4}$ (Rydberg decoherence) and $\Delta_{p,c,m}$ (peak shifts). In the following, we provide analytical solutions for $\rho_{ij}$ and $\Sigma_{ij}$ at low-probe-rate limit to facilitate understanding, while all fits and model simulations are carried out numerically by solving the master equations.

\paragraph{Three-level EIT}
%{\color{violet} The fit deph, in increasing order of probe rate, are :  $\Gamma_{33} = 0.13(4) , 0.17(5) , 0.44(6) , 0.90(8), 1.4(1)$ , $\Omega_c = 2.74(5)$}
We fit spectra from probe frequency scans in a three-level system (Fig.~\ref{fig: EIT nonlinearity models}) using Eqs.~\ref{eq:Liouvillian operator} and \ref{eq:H 2-level}, with no interaction-induced substitutions and $\Omega_m=0$. At the low-probe-rate limit, the numerical result asymptotically approaches 
\begin{equation}
    \rho_{21}=-\frac{i\lvert \Omega_p \rvert}{
(\Gamma_{2}-2i\Delta_p)
+\lvert \Omega_c \rvert^2/(\Gamma_{3}-2i\delta_2)}
\end{equation}
The fitted control Rabi frequency at the lowest probe rate is $\Omega_c/(2\pi)=2.74(5)$~MHz. This value of $\Omega_c$ is kept fixed when fitting the higher-rate data. The fitted $\Gamma_3$ at all rates are plotted in Fig.~\ref{fig: EIT nonlinearity}(c). The fitted probe resonance frequency is used to define their own $\Delta_p=0$ and shift each spectrum accordingly. In practice, all rates agree within 40~kHz. The fitted $-\Delta_c$ reflects the rate-dependent peak position. In Fig.~\ref{fig: EIT nonlinearity}(d), we instead determine the peak position by Gaussian fitting data within $\Delta_p/(2\pi)\in[-2,2]~$MHz. Peak positions determined in this manner statistically agree with the fitted $-\Delta_c$ while having smaller uncertainty. OD across all rates, except the highest one, agrees within 0.017. The highest rate data has an OD scaled by a factor of 0.76, which we attribute to atom loss from the probing region due to momentum kicks. 

In principle, probe Rabi frequency $\Omega_p$ can be estimated from the inferred probe rate $R_p$, together with the probe beam waist. In reality, this method can be inaccurate due to the uncertainty in the actual beam waist and its position relative to the atoms. If we enter the measured values, neither the linear nor the nonlinear models can capture the observed absorption baselines. Therefore, all fits and models in this paper use $\Omega_p$ values calculated from the measured probe beam waist ($1/e^2$ radius of 3.7~$\mu$m) and scaled down by a factor of 1.6, representing a global uncertainty likely associated with the effective probe size at the atom position. Notably, the control beam shows a similar scaling factor of 1.4 (calculated from its $1/e^2$ radius of 6.8~$\mu$m vs the fitted $\Omega_c$).
%confirming that this scaling factor represents a consistent systematic deviation for our beam waist measurements.
% Squared of waist enters rate. So scaling Rabi by a common factor is correct.
%\textbf{1.6 actually, I took 2um as the waist}

All models in Fig.~\ref{fig: EIT nonlinearity models} share the same input parameters from fitting the lowest-probe-rate data: $\Omega_c/(2\pi)=2.744$~MHz, $\Delta_c/(2\pi)=0.02$~MHz, OD = 0.404 (reduced to 0.309 for the highest-rate spectrum), and $\Omega_p/(2\pi)=0.0887~$MHz at the lowest rate (higher-rate model simulations scaled based on the measured probe rate). For the dephasing model, the interaction-induced $\Gamma_3/(2\pi)$ is added to a base value of 20~kHz. For conditional and unconditional models, $\Gamma_3/(2\pi)=20$ and 200~kHz, respectively. The number of atoms in each SA, $n_{\text{SA}} = \varrho V_b$, required by all models is derived from the atomic density $\varrho$, control Rabi frequency $\Omega_c$ and the van der Waals coefficient $C_6^{33}/(2\pi)=169\text{~GHz}~(\mu \text{m})^6$~\cite{vsibalic2017arc}. Given the parameters here, $n_{\text{SA}}=46$.
% Data shifted based on Delta_p=0, so simulation do not need to shift.

\paragraph{Four-level MW EIT}
\begin{figure*}[ht]
\centering
\includegraphics[width=0.9\textwidth]{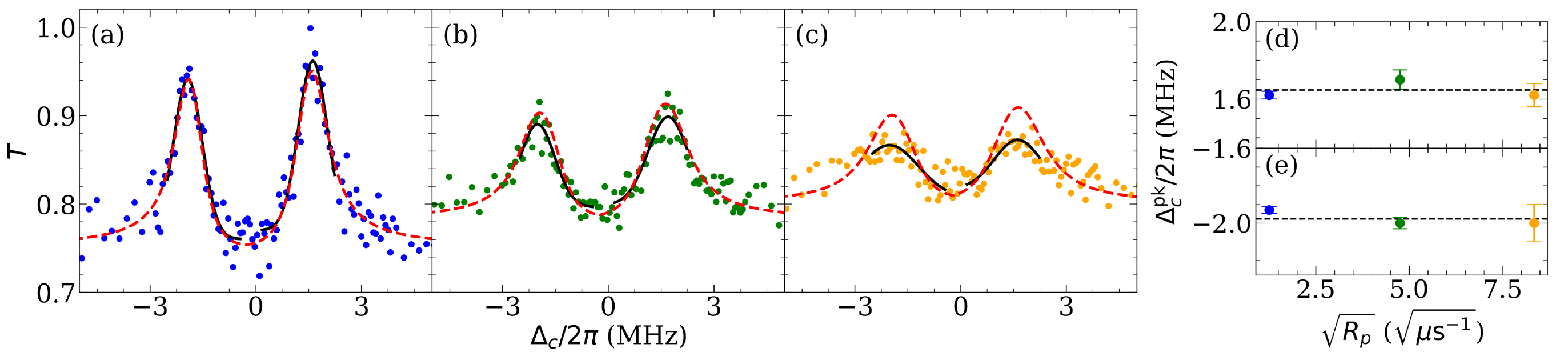}
\caption{Nonlinearity in a four-level system at larger probe Rabi frequency. Panels (a-c) are the measured MW EIT spectra vs control detuning with the probe rate $R_p=\{1.6,22.6,70.5\}~\mu\text{s}^{-1}$ for the \{blue, green, orange\} colors. $\Omega_c/(2\pi)=3.19(8)$~MHz and $\Omega_m/(2\pi)=3.45(3)$~MHz are extracted through the fit result of (a) from the linear model with independent $\Gamma_3$ and $\Gamma_4$. The red dashed lines are the dephasing model, and the black solid lines are Gaussian fits to each peak, from which the peak position $\Delta_c^{\text{pk}}$ are extracted and plotted in panel (d) (right peaks) and (e) (left peaks).}
\label{fig: 4level peak positions}
\end{figure*}

At a large bias magnetic field, we can isolate different four-level systems by varying the MW frequency to selectively bring the $\sigma^+$-, $\pi$-, or $\sigma^-$-polarized MW electric field component on resonance. 
By performing three separate spectroscopic measurements, each tuned to a different transition, and fitting the linear model, we extract $\{E_+,E_0,E_-\}=\{1.25(1),1.13(1),0.67(1)\}$~mV/cm, corresponding to
Rabi frequencies $\{\Omega_+,\Omega_0,\Omega_-\}/(2\pi)=\{2.04(2),2.61(3),1.88(3)\}~$MHz of $\ket{3_{-1/2}}\leftrightarrow\{\ket{4_{-3/2}},\ket{4_{-1/2}},\ket{4_{1/2}}\}$ transitions. The fit parameters are $\{\Delta_m,\Omega_m,\Omega_c,\Gamma_3,\Gamma_4,\text{OD}\}$, along with control resonance frequency. We set $\Delta_p=0$ from a three-level EIT probe-frequency scan in the low probe rate limit, where Rydberg-Rydberg interactions are negligible. The fitted control resonance frequency determines $\Delta_c=0$ and typically differs by less than $2\pi\times$0.2~MHz from its independently determined value from three-level EIT spectra. $\Omega_p$ is converted from probe rate in the same manner as for the three-level spectra.

%Alternatively, we can directly extract the electric field amplitude from the splitting of the peaks determined by Gaussian fitting. As such, we obtain $\{E_+,E_0,E_-\}=\{130(2),119(1),74(1)\}$~mV/cm. These values are systematically larger than the fitted values, which might be caused by insufficient splitting of the peaks as compared to their widths~\cite{holloway2017electric}. In both methods and all polarizations, we do not observe any statistically significant probe rate dependence of MW Rabi frequency.
%They correspond to Rabi frequencies $\{\Omega_+,\Omega_0,\Omega_-\}/(2\pi)=\{2.11(4),2.74(3),2.08(3)\}~$MHz of $\ket{3_{-1/2}}\leftrightarrow\{\ket{4_{-3/2}},\ket{4_{-1/2}},\ket{4_{1/2}}\}$ transitions.

The fitted control Rabi frequency at the lowest probe rate in Fig.~\ref{fig: EIT nonlinearity 2} is $\Omega_c/(2\pi)=2.76(6)$~MHz and is kept fixed when fitting data at higher probe rates.
%shows the $\sigma^+$-polarized MW EIT spectroscopy.
%Fit results of the lowest-rate data: $\Delta_m/(2\pi)=0.15(6)$, $\Omega_m/(2\pi)/(2\pi)=2.04(2)$, $\Omega_c/(2\pi)=2.76(6)$, $\Gamma_4/(2\pi)=0.21(7)$, $\Gamma_3/(2\pi)=0.17(8)$, OD = 0.347(4). 
$\{\Delta_m,\Omega_m,\Omega_c\}$ are input parameters to the dephasing and unconditional models presented in Fig.~\ref{fig:  dephasing spectrum}. 
The effective cloud length $L=64.8~\mu$m is kept fixed for all model simulations in this study, while the atomic number density $\varrho$, and therefore $n_{\text{SA}}$, is varied to account for different OD. Additionally, $R_b$ also varies slightly because different datasets have different $\Omega_c$.

Beyond the probe rate that shows a similar peak reduction to the three-level system, we measure at even higher probe rate (Fig.~\ref{fig: 4level peak positions}), up to the highest probe rate used in Fig.~\ref{fig: EIT nonlinearity}. We observe a similar loss of atoms, with a remaining fraction of 0.76 at the highest probe rate relative to the lowest probe rate, which is accounted for in the model by varying $\varrho$. The Gaussian-fit results show that we do not observe any spectral shift at this high rate within an uncertainty of 0.1~MHz. This supports that the shifts observed in Fig.~\ref{fig: EIT nonlinearity} are unlikely to be caused by effects associated with momentum kicks. 

%Among them, $E_+=xxx$ and $E_0=xxx$ agree with the multi-level MW EIT fit results, while $E_-=86(2)$ is fixed in the multi-level MW EIT fits. They correspond to Rabi frequencies $\{xxx,xxx,xxx\}~$MHz of $\ket{3_{-1/2}}\leftrightarrow\ket{4_{-3/2, -1/2, 1/2}}$ The same source power to the multilevel.

%The fit parameters in these spectra include OD, $\Omega_c$, $\Omega_m$, $\Delta_m$, and $\Gamma_3$. We determine the control resonance frequency and set $\Delta_p=0$ from a three-level EIT probe-frequency scan in the low probe rate limit, where Rydberg-Rydberg interactions are negligible. Such spectra also provide Rydberg decoherence rates typically ranging from $2\pi\times 0.1$ to $2\pi\times 0.2~$MHz. Based on this, we set $\Gamma_4$ to $2\pi\times0.2~$MHz. $\Omega_p$ are set to their estimated values; They are sufficiently low that they have an insignificant effect on the spectra. We present the four-level system data with $\ket{4_{1/2}}$ as the fourth level under three distinct probe Rabi frequencies in Fig.~\ref{fig: EIT nonlinearity 2}. The two peak positions are shown in panels (c,d), and the corresponding fitted $\Gamma_3$ values for $\Omega_p=2\pi\times(0.24,\,0.38,\,0.57)$~MHz are $2\pi\times(0.18(4),0.49(6),1.2(1))~$MHz.
% DC E field does not enter here, so $\Delta_m$ is not overfitting.

\subsection*{Interaction coefficients}
\begin{figure}[ht]
\centering
\includegraphics[width=0.46\textwidth]{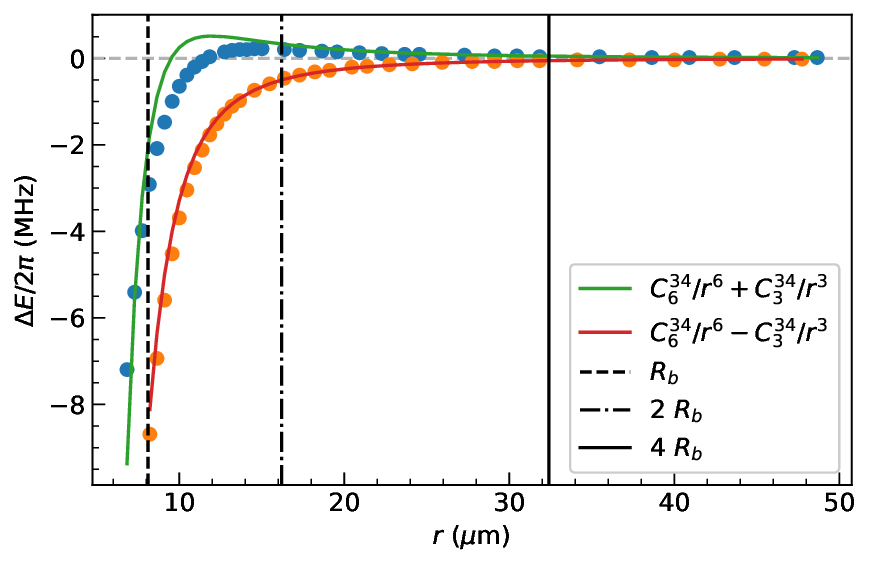}
\caption{ Interaction energy $\Delta E$ vs. interatomic distance $r$ for the $\ket{3_{-1/2}4_{1/2}}$ pairstate. $C_6^ {34}$ and $C_3^{34}$ are extracted from a simultaneous fit of the upper (blue) and the lower (orange) branches of the interaction potential obtained from the software \texttt{PairInteraction}~\cite{Weber2017}. The interatomic axis is aligned with the quantization axis ($\theta=0$) and B = 15.8 G.
%The dashed lines mark the start ($R_b$) and end (4$R_b$) points when we calculate the mean-fiedl shifts. {\color{violet}{Does the end point matter in the integration? Are we still effectively integrating to infinity, even though we imply the cutoff is important? No, I think end point is L/2.  L/2 is 4$R_b$. Then is there really any difference between right now and infinity? it matters to c3 not c6. Rb or 2Rb to 4Rb is only a tiny portion of c3.}}
}  
\label{fig: C6 and C3 fit}
\end{figure}

In addition to $C_6^{33}$, we also need $C_6^{44}/(2\pi)=-370\text{~GHz}~(\mu \text{m})^6$~\cite{vsibalic2017arc} and the interaction coefficients between levels $\ket{3_{-1/2}}$ and $\ket{4_{1/2}}$. As shown in Fig.~\ref{fig: C6 and C3 fit}, the pair potential exhibits both shifts and splitting. As a result, we jointly fit the two branches with the functions
$C_6^{34}/r^6 \pm C_3^{34}/r^3$, where the upper (lower) branch corresponds to the positive (negative) sign,
and obtain $C_6^{34}/(2\pi) = -1517\,\text{GHz}\,(\mu\text{m})^6$ and $C_3^{34}/(2\pi)=1.768\text{~GHz}~(\mu \text{m})^3$. $C_6^{44}$, $C_6^{34}$ and $C_3^{34}$ are all evaluated when the interatomic axis is aligned with the quantization axis, i.e., $\theta=0$.

%$C_3^{34}/(2\pi)=3.7\text{~GHz}~(\mu \text{m})^3$ for orientation of inter-atomic axis with respect to quantization axis equals to zero degree~\cite{vsibalic2017arc}. 

\subsection*{Model descriptions}
%Each SA contains $n_{SA}=\varrho V_b$ atoms, where $V_b=\frac{4\pi}{3}R_b^3$ is the blockade volume, and $\varrho$ is the atomic number density. Hence, 
We introduce the zero and first-order collective states $|1_{\mathrm{SA}}\rangle = |1_1, 1_2, \ldots, 1_{n_{\mathrm{SA}}}\rangle$, $|2_{\mathrm{SA}}\rangle = \frac{1}{\sqrt{n_{\mathrm{SA}}}}\sum_{j=1}^{{n_{\mathrm{SA}}}}\,|1_1, 1_2, \ldots, 2_j, \ldots, 1_{n_{\mathrm{SA}}}\rangle$, $|3_{\mathrm{SA}}\rangle = \frac{1}{\sqrt{n_{\mathrm{SA}}}}\sum_{j=1}^{{n_{\mathrm{SA}}}}\,|1_1, 1_2, \ldots, 3_j, \ldots, 1_{n_{\mathrm{SA}}}\rangle$, and $|4_{\mathrm{SA}}\rangle =\linebreak \frac{1}{\sqrt{n_{\mathrm{SA}}}}\sum_{j=1}^{{n_{\mathrm{SA}}}}\,|1_1, 1_2, \ldots, 4_j, \ldots, 1_{n_{\mathrm{SA}}}\rangle$. With these collective states as the basis states, we have $\varrho \rho_{33} \approx \varrho_{\mathrm{SA}} \Sigma_{33}$, $\varrho \rho_{44} \approx \varrho_{\mathrm{SA}} \Sigma_{44}$, and $\varrho \rho_{34} \approx \varrho_{\mathrm{SA}} \Sigma_{34}$, where $\varrho_{\mathrm{SA}}=V_b^{-1}$ is the SA number density. Assuming the MW on resonance, and the Rydberg decoherence is negligible, i.e. $\Delta_m=\Gamma_3=\Gamma_4=0$, we have
%\begin{widetext}
\begin{subequations}
\begin{align}
    \Sigma_{33}&
=
\frac{
n_{\mathrm{SA}} \lvert \Omega_c \rvert^2\lvert \tilde\Omega_p \rvert^2 \left( 2\delta_2 \right)^{2}
}{
\lvert A \rvert^{2}
+
n_{\mathrm{SA}} \lvert \Omega_c \rvert^2\lvert \tilde\Omega_p \rvert^2
\left[
\left( 2\delta_2 \right)^{2}
+
\lvert \Omega_m \rvert^2
\right]
}\\
    \Sigma_{44}&
=
\frac{
n_{\mathrm{SA}} \lvert \Omega_c \rvert^2\lvert \tilde\Omega_p \rvert^2 \lvert \Omega_m \rvert^2
}{
\lvert A \rvert^{2}
+
n_{\mathrm{SA}} \lvert \Omega_c \rvert^2\lvert \tilde\Omega_p \rvert^2
\left[
\left( 2\delta_2 \right)^{2}
+
\lvert \Omega_m \rvert^2
\right]
}\\
    \Sigma_{34}&
=
\frac{
n_{\mathrm{SA}} \lvert \Omega_c \rvert^2\lvert \tilde\Omega_p \rvert^2 \lvert \Omega_m \rvert(2\delta_2)
}{
\lvert A \rvert^{2}
+
n_{\mathrm{SA}} \lvert \Omega_c \rvert^2\lvert \tilde\Omega_p \rvert^2
\left[
\left( 2\delta_2 \right)^{2}
+
\lvert \Omega_m \rvert ^2
\right]
}%\\
%A&
%=
%(2\delta_2)
%\left[
%(2\Delta_p - i\Gamma_{2})(2\delta_2)
%-
%\lvert \Omega_c \rvert^2
%\right]
%-
%(2\Delta_p - i\Gamma_{2}) \lvert \Omega_m \rvert^2
\end{align}
\label{eq: state-state solution of coarse-grained solution}
\end{subequations}
%\end{widetext}
where $A
=
(2\delta_2)
\left[
(2\Delta_p - i\Gamma_{2})(2\delta_2)
-
\lvert \Omega_c \rvert^2
\right]
-
(2\Delta_p - i\Gamma_{2}) \lvert \Omega_m \rvert^2$. For all models, we obtain the three-level EIT results by setting $\Omega_m=0$.

The mean-field interactions are obtained by integrating over a spherical shell volume between radii $R$ and $R'$: 
%A realistic model of our atomic cloud would be uniform in the transverse plane and vary along the longitudinal (probe propagation) direction with a Gaussian profile. Nevertheless, it is common practice to model this density distribution as a homogeneous cloud with a different atomic density $\varrho$ and cloud length $L$, chosen such that the optical depth (OD) is preserved. We therefore assume a homogeneous atomic density within this cylindrical volume for the mean-field interactions:
%\begin{widetext}
\begin{subequations}
\begin{align}
        s_{33}&
=
\Sigma_{33}\varrho_{\mathrm{SA}}C_6^{33}
\int_{R<r<R'} d^3 r\,
\frac{1}{r^{6}}
\,  
%\exp\!\left(
%- \frac{(r\cos\theta)^2}{2\sigma^{2}}\right)
\\
        s_{44}&
=
\Sigma_{44}\varrho_{\mathrm{SA}} C_6^{44}
\int_{R<r<R'} d^3 r\,
\frac{\lvert 3\cos^{2}\theta - 1 \rvert}{2r^{6}}
\, 
%\exp\!\left(
%- \frac{(r\cos\theta)^2}{2\sigma^{2}}\right)
\\
        s_{34}&
=
\Sigma_{44}\varrho_{\mathrm{SA}}C_6^{34}
\int_{R<r<R'} d^3 r \,
\frac{\lvert 3\cos^{2}\theta - 1 \rvert}{2r^{6}}
\,  
\\
        s_{43}&
=
\Sigma_{33}\varrho_{\mathrm{SA}}C_6^{34}
\int_{R<r<R'} d^3 r\,
\frac{\lvert 3\cos^{2}\theta - 1 \rvert}{2r^{6}}
\,  
\\
        e_{34}&
=
|\Sigma_{34}|\varrho_{\mathrm{SA}} C_3^{34}
\int_{R<r<R'} d^3 r\,
\frac{\lvert 3\cos^{2}\theta - 1 \rvert}{2r^3} 
%\exp\!\left(
%- \frac{(r\cos\theta)^2}{2\sigma^{2}}\right)
\end{align}
\label{eq: The mean-field interactions}
\end{subequations}
%\end{widetext}
% Factor of 2 here because the original paper has all Rabi defined as half width.
where we have approximated the angular dependence of the anisotropic Rydberg–Rydberg interactions as $\frac{\lvert 3\cos^{2}\theta - 1 \rvert}{2}$.
%corresponding to the 
%44 and 34 pair states, characterized by the coefficients $C_6^{44}(\theta)\approx\frac{C_6^{44}(\theta=0)\,\lvert 3\cos^{2}\theta - 1 \rvert}{2}$ and $C_3^{34}(\theta)\approx\frac{C_3^{34}(\theta=0)\,\lvert 3\cos^{2}\theta - 1 \rvert}{2}$.

\paragraph{Conditional SA model}
%The original work~\cite{petrosyan2011electromagnetically} considers two separate contributions to the polarizability in a three-level EIT system: If an SA is excited, Rydberg blockade suppresses further Rydberg excitation within the blockade volume, so its polarizability reduces to that of an effective two-level atom without the control coupling. If the SA is not excited, surrounding SAs outside the blockade volume induce a mean-field interaction shift, which modifies the polarizability.

%When the interaction associated with either $C_6^{44}$ or $C_3^{34}$ is stronger than that arising from $C_6^{33}$ outside $R_b$, the model may require more fundamental changes than a naive extension. For example, the steady-state solution of the coarse-grained master equation (Eq.~\ref{eq: state-state solution of coarse-grained solution}) may require explicit inclusion of interaction effects. 

 The conditional probe-transition coherence $\rho_{21}$ is
\begin{equation}
    \rho_{21}= {\Sigma}_{33}\rho_{21}^{\text{2-level}}+(1-{\Sigma}_{33})\rho_{21}^{\text{MF}}
\end{equation}
with 
\begin{subequations}
    \begin{align}
    &\rho_{21}^{\text{2-level}}=-\frac{i\lvert \Omega_p \rvert}{\Gamma_{2}-2i\Delta_p}
    \\
   & \rho_{21}^{\text{MF}}=-\frac{i\lvert \Omega_p \rvert}{
(\Gamma_{2}-2i\Delta_p)
+\lvert \Omega_c \rvert^2/[\Gamma_{3}-2i(\delta_2-s_{33})]
}
\end{align}
\label{eq: mean-field rho21}
\end{subequations}
%where $\delta_3\equiv\Delta_c+\Delta_p+\Delta_m$ is the three-photon detuning, and $\Delta_m$ is the detuning of the MW. 

We consider 1D SA chain and divide the propagation length $L$ into $L/(2R_b)$ intervals. The input coherent probe field has local probe field intensity correlation function $g^{(2)}(z=0)=1$. At each SA, we evaluate $\Sigma_{33}(z)$ (Eq.~\ref{eq: state-state solution of coarse-grained solution}(a)) by weighting the local probe intensity $\lvert\Omega_p (z)\rvert^2$ with $g^{(2)}(z)$, i.e. $\lvert \tilde\Omega_p (z)\rvert^2\rightarrow \lvert \Omega_p (z)\rvert^2g^{(2)}(z)$. We perform Monte Carlo sampling accordingly: with probability $\Sigma_{33}$, $n_{33}=1$; and with probability $1-\Sigma_{33}$, $n_{33}=0$. To differentiate from the ensemble-averaged population $\Sigma_{33}$, we introduce a binary stochastic variable $n_{33}$ to represent the Monte Carlo outcome. The mean-field interactions (Eq.~\ref{eq: The mean-field interactions}) are evaluated by setting $g^{(2)}=1$. 
% We cannot plug in 1 or 0. If this is 1, it goes to the 2-level polarizability; if this is 0, the interactions are all zero. So clearly, we should not use the binary result.
% The paper does not put subscript r, so we don't weight this with g2.
The results are used to calculate
\begin{equation}
    \alpha(z)= n_{33}(z)\alpha^{\text{2-level}}+(1-n_{33}(z))\alpha^{\text{MF}}(z)
    \label{eq: polarizability}
\end{equation}
% $\alpha^{\text{MF}}(z)$ depends on z because $\lvert \Omega_p \rvert$ depends on z. Interaction depends on $\lvert \Omega_p \rvert$
where $\alpha$ denotes the polarizability, with $\alpha^{\text{2-level}}=-\rho_{21}^{\text{2-level}}\Gamma_2/\lvert \Omega_p \rvert$ and $\alpha^{\text{MF}}=-\rho_{21}^{\text{MF}}\Gamma_2/\lvert \Omega_p \rvert$. The following equations are used to determine $\lvert\Omega_p (z)\rvert^2$ and $g^{(2)}(z)$ for the next SA step, using the polarizability in Eq.~\ref{eq: polarizability}:
%\begin{widetext}
\begin{subequations}
    \begin{align}
        \partial_z (\lvert \Omega_p \rvert^2)
&=
- \kappa(z)\,
\lvert \Omega_p \rvert^2\,\mathrm{Im}[\alpha(z)]\\
\partial_z g_p^{(2)}(z)
&=
- \kappa(z)\,
n_{33}(z)
\,\mathrm{Im}\!\left[ \alpha^{\text{2-level}}- \alpha^{\text{MF}}(z) \right]
\, g_p^{(2)}(z)
    \end{align}
\end{subequations}
%\end{widetext}
where $ \kappa(z)=2\varrho\omega_p \lvert \wp_{12} \rvert^{2} / (\hbar \epsilon_0 c \Gamma_2)$. Here, $\wp_{12}$ is the dipole moment of the $\ket{1}\leftrightarrow\ket{2}$ transition, and $\omega_p$ is the corresponding transition angular frequency.

\paragraph{Unconditional SA model}
In this model, $\rho_{21}= \rho_{21}^{\text{MF}}$ unconditionally. For the three-level EIT system, it takes the same form as Eq.~\ref{eq: mean-field rho21}(b); for the four-level MW EIT system, it is 
\begin{align}
    \rho_{21}&=-\frac{i\lvert \Omega_p \rvert}{
(\Gamma_{2}-2i\Delta_p)
+\frac{\lvert \Omega_c \rvert^2
(\Gamma_{4}-2i\delta_3')}
{\lvert \Omega_m+2e_{34} \rvert^2
+
(\Gamma_{3}-2i\delta_2')
(\Gamma_{4}-2i\delta_3')}}
\end{align}
with $\delta_2'=\Delta_p+\Delta_c-s_{33}-s_{34}$ and $\delta_3'=\Delta_p+\Delta_c+\Delta_m-s_{44}-s_{43}$.
When modeling the four-level system, because of the slow spatial decay of the $C_3^{34}$ interaction, in Eq.~\ref{eq: The mean-field interactions} we set the upper limit of the integration range to $R'=L/2$. Due to the differences in the interaction strengths, each integral has its own lower cutoff $R$, set by equating the corresponding interaction to $\Omega_c^2/(2\Gamma_2)$. $\Gamma_{3,4}/(2\pi)=0.2~$MHz is included in the model to match our measurements at the lowest probe rates.
%The original work~\cite{wang2025enhanced} suggests integrating starting from the interatomic distance $R=\varrho^{-1/3}$ to obtain the mean-field interactions. This leads to enormous amount of spectral peak shifts which we do not observe in either our three-level or four-level systems. Following the spirit of SAs, we instead integrating 

% \textbf{For $\bm{\Omega_c = 2.744}$ (Fig2 value), $\bm{R_b = 8.06}$, for $\bm{\Omega_c = 2.76}$ (Fig3 value), $\bm{R_b = 8.04}$, $\bm{R_{b,44} = 9.17}$, $\bm{R_{b,34,s} = 11.6}$, $\bm{R_{b,34,e} = 14.2}$, for $\bm{\Omega_c = 3.27}$ (Fig 5 value), $\bm{R_b = 7.6}$, $\bm{R_{b,44} = 8.7}$, $\bm{R_{b,34,s} = 11.0}$, $\bm{R_{b,34,e} = 12.7}$}

\paragraph{Dephasing model}
In this model, the effects of interactions are included as increased decoherence rather than shifts:
\begin{align}
    \rho_{21}&=-\frac{i\lvert \Omega_p \rvert}{
(\Gamma_{2}-2i\Delta_p)
+\frac{\lvert \Omega_c \rvert^2
(\Gamma_{4}'-2i\delta_3)}
{\lvert \Omega_m \rvert^2
+
(\Gamma_{3}'-2i\delta_2)
(\Gamma_{4}'-2i\delta_3)}}
\end{align}
%This equation is obtained by setting $s_{33}=s_{44}=e_{34}=0$ in Eq.~\ref{eq: mean-field rho21}(b), while replacing $\Gamma_3$ with \textcolor{blue}
with $\Gamma_3'=2\pi\times20~\text{kHz}+\sqrt{\text{Var}_{33}}+\sqrt{\text{Var}^s_{34}}+\sqrt{\text{Var}^e_{34}}$, $\Gamma_4'=2\pi\times20~\text{kHz}+\sqrt{\text{Var}_{44}}+\sqrt{\text{Var}^s_{43}}+\sqrt{\text{Var}^e_{34}}$, and
%{$\Gamma_3'=\Gamma_3+\sqrt{\text{Var}_{33}}+\sqrt{\text{Var}_{44}}+2\sqrt{\text{Var}_{34}}$. ?}
\begin{subequations}
    \begin{align}
         \text{Var}_{33}&
=
\Sigma_{33}\varrho_{\mathrm{SA}}
\!\!\int_{R<r<R'} \!\!\!\!\!\!d^3 r
\left[\frac{C_6^{33}}{r^{6}}-\overline{\frac{C_6^{33}}{r^{6}}}\right]^2
\\ 
\text{Var}_{44}&
=
\Sigma_{44}\varrho_{\mathrm{SA}} \!\!
\int_{R<r<R'} \!\!\!\!\!\!d^3 r
\left[\frac{C_6^{44}(\theta)}{r^{6}}-\overline{\frac{C_6^{44}(\theta)}{r^{6}}}\right]^2
\\
\text{Var}^s_{34}&
=
\Sigma_{44} \varrho_{\mathrm{SA}} \!\!
\int_{R<r<R'} \!\!\!\!\!\!d^3 r
\left[\frac{C_6^{34}(\theta)}{r^{6}}-\overline{\frac{C_6^{34}(\theta)}{r^{6}}}\right]^2
\\
\text{Var}^s_{43}&
=
\Sigma_{33} \varrho_{\mathrm{SA}} \!\!
\int_{R<r<R'} \!\!\!\!\!\!d^3 r
\left[\frac{C_6^{34}(\theta)}{r^{6}}-\overline{\frac{C_6^{34}(\theta)}{r^{6}}}\right]^2
\\
\text{Var}^e_{34}&
=
|\Sigma_{34}| \varrho_{\mathrm{SA}} \!\!
\int_{R<r<R'} \!\!\!\!\!\!d^3 r
\!\left[\frac{C_3^{34}(\theta)}{r^{3}}-\overline{\frac{C_3^{34}(\theta)}{r^{3}}}\right]^2
    \end{align}
\end{subequations}
where
%\begin{widetext}
\begin{subequations}
    \begin{align}    \overline{\frac{C_6^{33}}{r^{6}}}&=\frac{1}{\mathcal{N}}
\int_{R<r<R'} d^3 r
\frac{C_6^{33}}{r^{6}}
%\, \exp\!\left(
%- \frac{(r\cos\theta)^2}{2\sigma^{2}}\right)
\\
\overline{\frac{C_6^{ij}(\theta)}{r^{6}}}&=\frac{1}{\mathcal{N}}
\int_{R<r<R'} d^3 r
\frac{C_6^{ij}(\theta)}{r^{6}}
%\, \exp\!\left(- \frac{(r\cos\theta)^2}{2\sigma^{2}}\right)
\\
         \overline{\frac{C_3^{34}(\theta)}{r^{3}}}&=\frac{1}{\mathcal{N}}
\int_{R<r<R'} d^3 r
\frac{C_3^{34}(\theta)}{r^{3}}
%\, \exp\!\left(- \frac{(r\cos\theta)^2}{2\sigma^{2}}\right)
    \end{align}
\end{subequations}
%\end{widetext}
with $C_6^{ij}(\theta)=C_6^{ij}\frac{\lvert 3\cos^{2}\theta - 1 \rvert}{2}$, $C_3^{34}(\theta)=C_3^{34}\frac{\lvert 3\cos^{2}\theta - 1 \rvert}{2}$, and the normalization factor $\mathcal{N}=
\int_{R<r<R'} d^3 r =\frac{4\pi}{3}(R'^3-R^3)$. The integration range cutoffs $R$ and $R'$ are set in the same manner as for the unconditional model. Additionally, $\tilde{\Omega}_p=\Omega_p$ for both models.

\section*{Data availability}
The data and code that support the findings of this study are openly available in Zenodo at https://doi.org/10.5281/zenodo.18560777.
%\bibliographystyle{apsrev4-2}
%\bibliography{references}

\section*{Acknowledgments}
We thank Profs. Chen-Lung Hung and 	
Jonathan D. Hood for loaning us equipment. We are grateful for valuable discussions with Chen-Lung Hung. This work was supported by Purdue startup fund and AFOSR Grant FA9550-22-1-0327.

\clearpage
\onecolumngrid

\begin{center}
\textbf{\large Supplementary Information:}\\
\textbf{\large Nonlinear optical spectra from Rydberg-mediated photon-photon interactions}
\end{center}

\setcounter{section}{0}
\renewcommand{\thesection}{S\arabic{section}}
\renewcommand{\thefigure}{S\arabic{figure}}
\renewcommand{\thetable}{S\arabic{table}}

\section*{Part 1. Discrepancy Between AT-Splitting and Fitted Microwave Rabi Frequency}

In this section, we report the discrepancy between the given microwave Rabi frequency $\Omega_m$ and the corresponding Autler–Townes (AT) splitting $\Delta_{\text{AT}}$, which is extracted from the peak positions of the two maxima in the simulated spectra for different values of $\Gamma_3$ and $\Gamma_4$. The fixed parameters used in the simulations are $\{\Delta_m,\Omega_m,\Omega_c\}/(2\pi)=\{0,2.04,2.76\}$~MHz and $\mathrm{OD}=0.347$, taken from the fit results of Fig.~3(e) in the main text, except that $\Delta_m$ is set to zero. The spectra are simulated over the ranges $\Delta_c/(2\pi)\in[-\Omega_m/(4\pi)-0.5,-\Omega_m/(4\pi)+0.5]$~MHz and $\Delta_c/(2\pi)\in[\Omega_m/(4\pi)-0.5,\Omega_m/(4\pi)+0.5]$~MHz, using 2000 sampling points for each interval, corresponding to a frequency resolution of 0.5~kHz.

%The spectra are simulated over the ranges $\Delta_c/(2\pi)\in[-\Omega_m-0.5,-\Omega_m+0.5]$~MHz and $\Delta_c/(2\pi)\in[\Omega_m-0.5,\Omega_m+0.5]$~MHz, using 2000 sampling points for each interval, corresponding to a frequency resolution of 0.5~kHz.

\begin{figure*}[ht]
\centering
\includegraphics[width=1.0\textwidth]{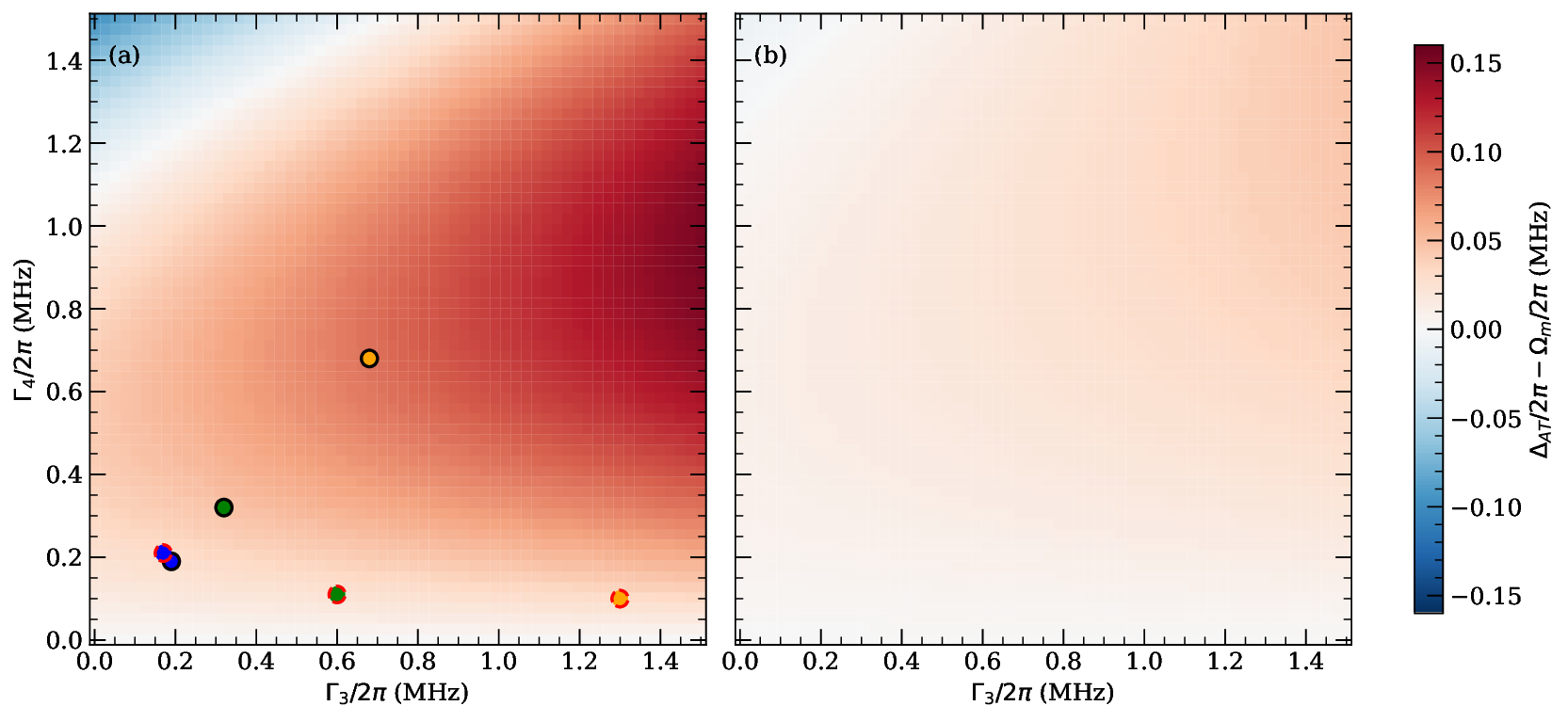}
\caption{AT-splitting and curvefit result discrepancy. (a) Calculated discrepancy with the fit results of the data in Fig.~3(e) in the main text except $\Delta_m$ is set to be 0 as a function of $\Gamma_3$ and $\Gamma_4$. Colored markers correspond to the three representative parameter sets with the red dashed and black solid marker edges denote the two fitting strategies within the same linear model, as shown in Fig.~3(e)–(g) of the main text. (b) Same calculation as in (a) but with the $\Omega_m$ increased by a factor of five.}
\label{fig:  discrepancy}
\end{figure*}

Fig.~\ref{fig: discrepancy}(a) clearly shows that, within our operating regime, a systematic discrepancy exists between the $\Delta_\text{AT}$ the $\Omega_m$ used in the model. The simulated discrepancy ($(\Delta_\text{AT}-\Omega_m)/(2\pi)$) is $+0.04$~MHz, which agrees with the value reported in the main text ($+0.07(4)$~MHz) within the experimental uncertainty. In contrast, when the microwave Rabi frequency is increased by a factor of five (Fig.~\ref{fig:  discrepancy}(b)), the peak separation becomes much larger than the linewidth. In this regime the two resonances are well resolved, the lineshape approaches the ideal AT picture, and the discrepancy vanishes.

\newpage

\section*{Part 2. Fit Results for Fig. 5 and Additional Transitions}

\begin{figure*}[ht]
\centering
\includegraphics[width=0.7\textwidth]{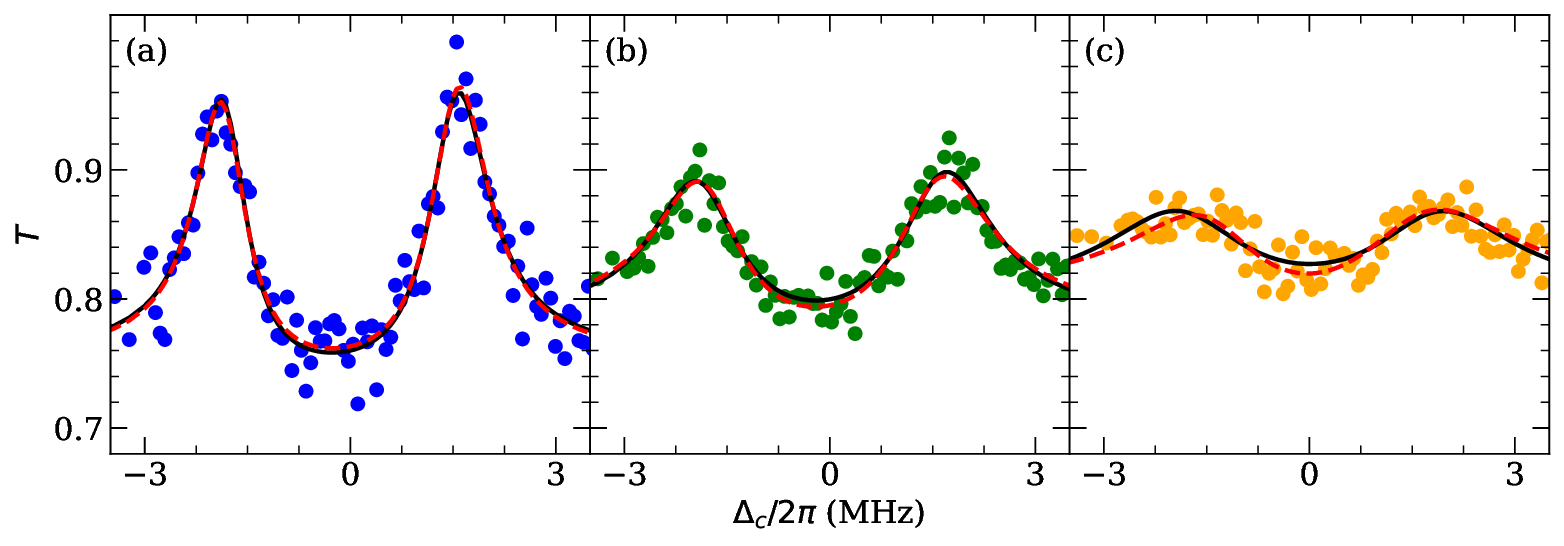}
\caption{Fit results from the linear model for the data shown in Fig.~5 of the main text. In panels (a-c), the red dashed lines have independent $\Gamma_3$ and $\Gamma_4$, with fitted values $\Gamma_3/(2\pi)=\{0.0(1),1.1(2),2.5(3)\}~$MHz and $\Gamma_4/(2\pi)=\{0.3(1),0.4(1),0.8(2)\}$~MHz. The black solid lines enforces $\Gamma_3=\Gamma_4$, with fitted values $\Gamma_{3,4}/(2\pi)\ = \{0.16(3),0.74(5),1.6(1)\}$~MHz.}
\label{fig:  Fig5 fit result}
\end{figure*}

\begin{figure*}[ht]
\centering
\includegraphics[width=0.7\textwidth]{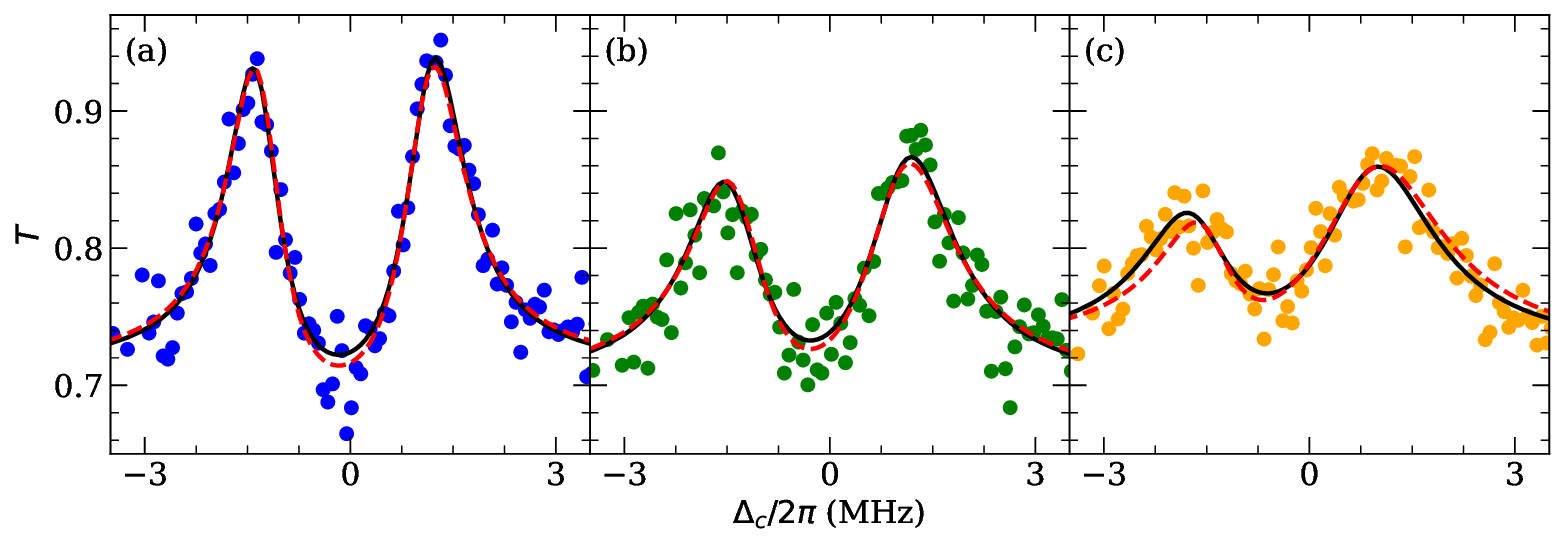}
\caption{Fit results from the linear model for the $\pi$ transition data. In panels (a-c), the red dashed lines have independent $\Gamma_3$ and $\Gamma_4$, with fitted values $\Gamma_3/(2\pi)=\{0.4(1),0.9(2),1.2(1)\}~$MHz and $\Gamma_4/(2\pi)=\{0.04(8),0.5(1),0.7(2)\}$~MHz. The black solid lines enforces $\Gamma_3=\Gamma_4$, with fitted values $\Gamma_{3,4}/(2\pi)\ = \{0.21(3),0.66(5),1.00(6)\}$~MHz.}
\label{fig:  pi fit result}
\end{figure*}

\begin{figure*}[ht]
\centering
\includegraphics[width=0.51\textwidth]{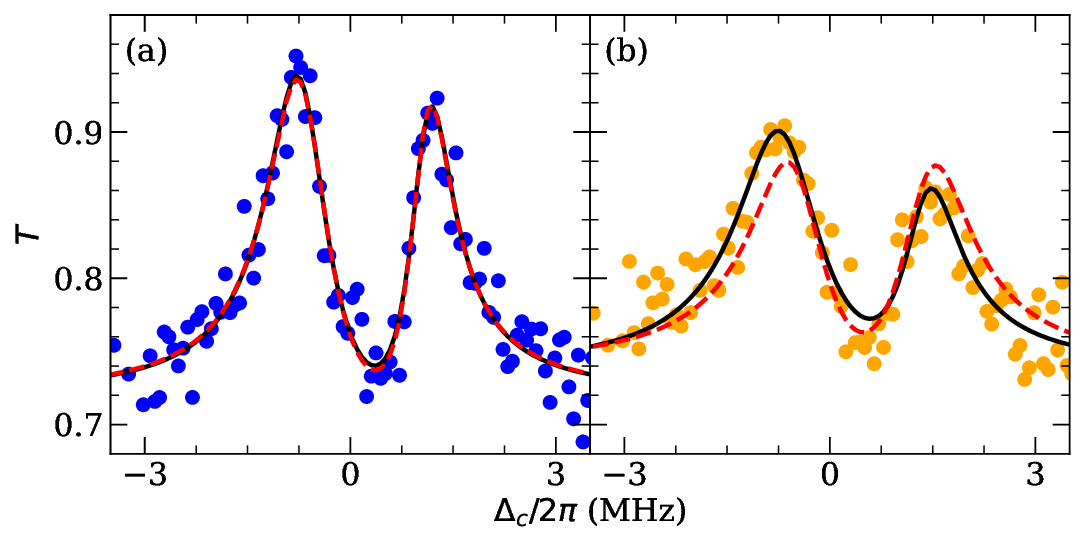}
\caption{Fit results from the linear model for  the $\sigma^-$ transition data. In panels (a-b), the red dashed lines have independent $\Gamma_3$ and $\Gamma_4$, with fitted values $\Gamma_3/(2\pi)=\{0.27(7),0.8(2)\}~$MHz and $\Gamma_4/(2\pi)=\{0.18(6),0.3(1)\}$~MHz. The black solid lines enforces $\Gamma_3=\Gamma_4$, with fitted values $\Gamma_{3,4}/(2\pi)\ = \{0.22(2), 0.54(4)\}$~MHz.}
\label{fig:  sigma- fit result}
\end{figure*}

\bibliography{references}

\end{document}